\documentclass[%
 aip,
amsmath,amssymb,
reprint,
]{revtex4-1}

\usepackage{graphicx}
\usepackage{dcolumn}
\usepackage{bm}

\usepackage[utf8]{inputenc}
\usepackage[T1]{fontenc}
\usepackage{mathptmx}
\usepackage{etoolbox}
\usepackage{physics}
\usepackage{tikz}
\usepackage{mathdesign}
\usepackage{xcolor}

\newcommand{\R}{\bm R}
\newcommand{\B}{\bm B}
\newcommand{\E}{\bm E}
\newcommand{\A}{\bm A}
\newcommand{\J}{\mathcal{J}}
\newcommand{\N}{\mathcal{N}}
\newcommand{\vi}{\bm v}

\newcommand{\C}{ \mathcal{C}}

\renewcommand{\|}{\parallel}

\newcommand{\projpj}[2]{ \norm{#2}^{#1}}

\renewcommand{\poissonbracket}[2]{  \left[#1, #2 \right]}

\renewcommand{\eqref}[1]{Eq.~(\ref{#1})}

\newcommand{\figref}[1]{Fig.~\ref{#1}}
\newcommand{\Figref}[1]{Figure.~\ref{#1}}
\newcommand{\secref}[1]{Sec.~\ref{#1}}

\makeatletter
\def\@email#1#2{%
 \endgroup
 \patchcmd{\titleblock@produce}
  {\frontmatter@RRAPformat}
  {\frontmatter@RRAPformat{\produce@RRAP{*#1\href{mailto:#2}{#2}}}\frontmatter@RRAPformat}
  {}{}
}%
\makeatother
\begin{document}

\preprint{AIP/123-QED}

\title[]{Full-F Turbulent Simulation in a Linear Plasma Device using a Gyro-Moment Approach}

\author{B. J. Frei}
 \email{baptiste.frei@ipp.mpg.de}
  \affiliation{Ecole Polytechnique F\'ed\'erale de Lausanne (EPFL), Swiss
Plasma Center, CH-1015 Lausanne, Switzerland
}
 \affiliation{Max-Planck-Institut für Plasmaphysik, D-85748 Garching, Germany
}

\author{J. Mencke}
\affiliation{Ecole Polytechnique F\'ed\'erale de Lausanne (EPFL), Swiss
Plasma Center, CH-1015 Lausanne, Switzerland
}

\author{P. Ricci}
 \affiliation{Ecole Polytechnique F\'ed\'erale de Lausanne (EPFL), Swiss
Plasma Center, CH-1015 Lausanne, Switzerland
}

\date{\today}
\begin{abstract}
Simulations of plasma turbulence in a linear plasma device configuration are presented. These simulations are based on a simplified version of the gyrokinetic (GK) model proposed by B. J. Frei \textit{et al.} [J. Plasma Phys. \textbf{86}, 905860205 (2020)] where the full-F distribution function is expanded on a velocity-space polynomial basis allowing us to reduce its evolution to the solution of an arbitrary number of fluid-like equations for the expansion coefficients, denoted as the gyro-moments (GM). By focusing on the electrostatic and neglecting finite Larmor radius effects, a full-F GM hierarchy equation is derived to evolve the ion dynamics, which includes a nonlinear Dougherty collision operator, localized sources, and Bohm sheath boundary conditions. An electron fluid Braginskii model is used to evolve the electron dynamics, coupled to the full-F ion GM hierarchy equation via a vorticity equation where the Boussinesq approximation is used. A set of full-F turbulent simulations are then performed using the parameters of the LArge Plasma Device (LAPD) experiments with different numbers of ion GMs and different values of collisionality. The ion distribution function is analyzed illustrating the convergence properties of the GM approach. In particular, we show that higher-order GMs are damped by collisions in the high-collisional regime relevant to LAPD experiments. The GM results are then compared with those from two-fluid Braginskii simulations, finding qualitative agreement in the time-averaged profiles and statistical turbulent properties.
\end{abstract}

\maketitle

\section{Introduction}

Despite recent progress in the development of gyrokinetic (GK) codes, such as \verb|COGENT| \cite{Dorf2020}, \verb|Gkeyll| \cite{Hakim2020}, \verb|GENE-X| \cite{Michels2022,Ulbl2023} and \verb|XGC| \cite{Chang2017}, extending the GK model from the core to the boundary remains challenging since it requires dealing with a wide range of collisionality, order-one fluctuations across various scales, complex magnetic field geometry, steep pressure gradients and the interaction of the plasma with the wall. As a consequence, less computationally demanding tools such as fluid simulations (see, e.g., Refs. \onlinecite{Stegmeir2018,De2022,Giacomin2022}) based on the drift-reduced Braginskii model \cite{Zeiler1997}, are often used to simulate the plasma dynamics in the boundary. However, the Braginskii fluid approach remains limited to the highly collisional region of the boundary, namely the scrape-off layer (SOL) and cannot describe kinetic effects at lower collisionality.

To tackle the challenges of the boundary region, an approach is formulated in Ref. \onlinecite{Frei2020} based on the Hermite-Laguerre expansion of the full (full-F) distribution function, which leads to the evolution of an arbitrary number of expansion coefficients, referred to as the gyro-moments GMs. This approach features kinetic effects \cite{Jorge2019epw,Frei2022,Frei2023}, which are absent in Braginskii-like fluid models, and collisional effects modeled using advanced collision operators \cite{Jorge2019,Frei2021,Frei2022b,Frei2023,Hoffmann2023} with an arbitrary number of GMs. The ability to adjust the number of GMs with a simple closure by truncation removes the need for \textit{ad-hoc} kinetic closures \citep{Hammett1990,Beer1996} since the accuracy of the model can be improved by increasing arbitrarily the number of GMs and, therefore, the velocity-space resolution. So far, investigations based on the GM approach are limited to the $\delta f$ regime (where only the \textit{a priori} small deviation of the distribution function from thermal equilibrium is evolved \citep{Brizard2007}) showing convergence with a low number of GMs, in particular at high collisionality \cite{Frei2023}. We notice that the first nonlinear $\delta f$ simulations using advanced linearized collision operators have been recently performed \cite{Hoffmann2023,Hoffmann2023b} and that a similar approach has also been implemented to perform $\delta f$ turbulent calculations focusing on the core region \cite{Mandell2018,Mandell2022}. In this work, we present the first full-F turbulent results that is based on the GM approach using a flexible number of moments. In particular, we focus on simulations of plasma turbulence in a linear plasma device.

Linear plasma devices, such as LAPD \cite{Gekelman1991}, HelCat \cite{Gilmore2015} and RAID \cite{Furno2017}, are experiments that allow for the investigation of basic plasma phenomena in a simplified magnetic geometry characterized by the absence of magnetic gradients, curvature, and shear \cite{Gekelman1991,Tynan2004,Castellanos2005,Rogers2010,Furno2017}. Despite their simplicity and the lack of kinetic effects such as trapped electrons, linear plasma devices share some of the most important physical processes that occur in the boundary of magnetic confinement devices. In fact, similar to the boundary, the turbulent dynamics in a linear plasma device result from the interplay of cross-field transport, parallel flows to the magnetic field, and plasma losses at the end plates where a sheath forms due to plasma-wall interactions. At the same time, the straight magnetic field lines in these devices facilitate the development of new modelling tools, compared to complex magnetic geometry characterizing the boundary of fusion devices. The modelling in these devices is also simplified by the perpendicular incidence of the magnetic field lines to the wall of the machine, which simplifies the sheath boundary model compared to an oblique incidence \cite{Chodura1982,Loizu2012,Geraldini2017} and by the low plasma temperatures comparable to typical SOL values (e.g., $ T_i \lesssim  T_e \sim 6 $ eV in typical LAPD discharges \cite{Friedman2012}), which are ideal for applying the full-F GM approach.

The low plasma temperature allows for a direct comparison of the GM approach with validated fluid simulations \cite{Rogers2010, Popovich2010,Fisher2015,Ross2019}, which have been considered because of the collisional conditions often met in, e.g., LAPD experiments. Moreover, we note that, in addition to fluid simulations, LAPD configuration was also chosen to perform the first full-F GK simulation in open field lines with the \verb|Gkeyll| code that uses a discontinuous-Galerkin approach to discretize the velocity-space \cite{Shi2017}. LAPD turbulent simulations using the GK \verb|GENE| code are also reported in Ref. \onlinecite{Pan2018} based on the same physical model. In both cases, the GK model used for the LAPD simulations neglects finite Larmor radius (FLR) effects and can, therefore, be considered equivalent to a drift-kinetic model. As a matter of fact, previous fluid and GK simulations of LAPD provide a comparison that make linear plasma devices an ideal testbed to perform the first full-F turbulent simulations using the GM approach with an arbitrary number of GMs.

In this work, we consider a simplified version of the full-F GM model derived in Ref. \onlinecite{Frei2020}. Consistently with Refs. \onlinecite{Shi2017} and \onlinecite{Pan2018}, we focus on the long-wavelength electrostatic limit of the GK model to describe the ion dynamics, with ion-ion collisions modeled using a simple nonlinear Dougherty \cite{Dougherty1964} collision operator (similar to the one used in Refs. \onlinecite{Shi2017,Pan2018}). In addition, we neglect FLR effects and we refer to our ion model as a GK model for consistency with previous works \cite{Shi2017,Pan2018}, but emphasize its similarity to a drift-kinetic ion model. On the other hand, electrons are assumed collisional, such that their dynamics can be approximated by the drift-reduced Braginskii model \cite{Zeiler1997}. This is in contrast with respect to previous GK simulations where electrons are treated using the long-wavelength (without FLR effects) GK equation. Hence, our model can be considered as a hybrid kinetic-fluid model, with a GK ion and fluid electron description.

In contrast to previous GK simulations of linear devices \cite{Shi2017,Pan2018}, the ion GK equation is solved within the GM approach where the full ion distribution function $F_i$ is expanded on a Hermite and Laguerre polynomial basis. A parallel (to the magnetic field) velocity-space coordinate shifted by the local ion parallel fluid velocity and the adiabatic invariant are used to describe efficiently sonic ion parallel flows near the end plates where the sheath forms. A full-F ion GM hierarchy equation for the expansion coefficients is then derived. The ion full-F GM hierarchy equation and the fluid electron model are coupled through a vorticity equation where the Boussinesq approximation is considered. To incorporate the losses at the end plates, Bohm sheath boundary conditions \cite{Loizu2012} are implemented in the parallel direction, which are equivalent to the ones used in the previous Braginskii simulation of LAPD \cite{Rogers2010,Fisher2015}. Nonlinear simulations of LAPD are then performed with various numbers of GMs. For comparison, a set of nonlinear turbulent simulations are also performed using the two-fluid drift-reduced Braginskii equations \cite{Zeiler1997} (or simply Braginskii model), similarly to Refs. \onlinecite{Rogers2010,Fisher2015}, and using a reduced cold-ion model derived from the full-F ion GM hierarchy.  

The present results demonstrate that the full-F GM approach properly describes fluctuations in an open-field line geometry. In fact, a detailed analysis shows that turbulence, driven by a long perpendicular wavelength Kelvin-Helmholtz (KH) instability, is in qualitative agreement with the Braginskii model. The importance of the KH instability in determining the radial turbulent transport in LAPD, as pointed out by Ref. \onlinecite{Rogers2010}, motivate the long-wavelength approximation and neglecting FLR effects in the present work. Our findings exhibit weak dependence on the number of GMs employed in the simulations and on the collisional regime. This is primarily due to the absence of significant kinetic effects in LAPD and the fluid nature of the KH instability, which governs radial turbulent transport and manifests itself at long perpendicular wavelengths. The analysis of the velocity-space representation of the ion distribution function demonstrates that the amplitude of the GMs decays rapidly with the order of the polynomial when collisions are considered. The present investigation also reveals that a simple closure based on the truncation of the GM hierarchy is sufficient in our case and has little effect on turbulence. It is important to note that the purpose of these simulations is not to achieve a highly-fidelity and realistic description of LAPD turbulence, but rather to establish confidence in the applicability of the GM approach in full-F turbulent calculations with a flexible number of GMs. Furthermore, a direct comparison with LAPD experimental data \cite{Carter2006,Fisher2015} and with previous GK simulations \cite{Shi2017, Pan2018} falls outside the scope of our study, but will be addressed in future work.

We remark that the GM approach of Ref. \onlinecite{Frei2020} shares some similarities with previous (full-F or $\delta f$) gyrofluid models \cite{Brizard1992,Beer1996,Snyder2001,Strintzi2004,Madsen2013,Held2016,Wiesenberger2019,Galassi2022,Wiesenberger2022,Wiesenberger2023}, where a fixed number of moments, usually up to the third-order velocity-space moments, are evolved as dynamical variables. These models can be considered as the limit of the GM approach where a finite number of moments is evolved and properly designed closures to mimic kinetic effects, such as Landau damping and FLR effects. Hence, the GM approach used in this work is similar to a full-F gyrofluid model without FLR effects when the number of moments is kept constant. Finally, although our model considers a closure by truncation, FLR closures at arbitrary perpendicular wavelengths \cite{Held2020,Held2022,Wiesenberger2023} and collisional closures \cite{Jorge2017,Wiesenberger2022} for full-F gyrofluid models can also be considered with an arbitrary number of GMs.

The paper is structured as follows. In \secref{sec:linearplasmadevicemodel}, we derive the ion full-F GM hierarchy equation in a straight magnetic field and introduce the electron fluid model, as well as the two-fluid drift-reduced Braginskii model. The numerical implementation of the full-F GM hierarchy equation is detailed in \secref{sec:numericalimplementation}. The results of the first full-F GM turbulent simulations are presented in \secref{sec:turbulentsimulations}, which includes a detailed comparison with the Braginskii simulations and an analysis of the ion distribution function. We conclude in \secref{sec:conclusion}. Appendix \ref{appendixA} reports on the derivation of the ion temperature equation at high collisionality using the GM approach.

\section{Linear Plasma Device Model}
\label{sec:linearplasmadevicemodel}

In this section, we derive the full-F GM hierarchy equation for the ion dynamics by expanding the ion distribution function onto a Hermite and Laguerre polynomials basis. The hierarchy includes particle and energy sources and a simple nonlinear long-wavelength Dougherty collision operator. The ion full-F GM hierarchy is derived by neglecting FLR effects from gyro-average and the difference between particle and gyrocenter fluid quantities, but includes long-wavelength approximation of the polarization density in the quasineutrality equation where the Boussinesq approximation is considered. Ultimately, this allows us to remove the complexities of FLR effects in full-F calculations using a flexible number of GMs. A reduced cold-ion model is also considered, for comparison purposes, which is obtained analytically from the full-F GM hierarchy in the cold-ion ($T_i =0$) limit. For the electron dynamics, the Braginskii fluid equations are used to evolve the electron density $n_e$, parallel velocity $U_{\| e}$, and temperature $T_e$. A vorticity equation is derived using the Boussinesq approximation for the electrostatic potential $\phi$, which couples the ion and electron models. Finally, we present the two-fluid Braginskii model.

The simple magnetic geometry in a linear plasma device allows us to introduce a simple coordinate system. In particular, assuming a rectangular shape of the linear device cross-section, we define the cartesian coordinate system $(x,y,z)$, such that the $(x,y) $ coordinates describe the plane perpendicular to $\B$, while $z$ is the coordinate along the magnetic field lines. The height and width of the perpendicular cross-section are $L_x$ and $L_y$, respectively, and the length of the linear plasma device is $L_z$. The constant in time and uniform in space magnetic field can simply be written as $\B =  \grad \times \bm A = B \bm z$, where $\bm A$ is the magnetic vector potential, which is constant in time, and $\bm z$ is the unit vector pointing along the axis of the linear device.

 This section is structured as follows. We describe the ion full-F model in \secref{subsec:iongkmodel} and we derive the ion full-F GM hierarchy equation in \secref{subsec:fullFhierarchy}. A presentation of the reduced cold-ion model is then obtained from the GM hierarchy equation in \secref{subsec:coldion}. The fluid electron model follows in \secref{subsec:electronbraginskii} and the vorticity equation is derived in \secref{subsec:vorticityequation}. \secref{subsec:braginskii} describes the two-fluid Braginskii model used for comparison purposes and, finally, \secref{subsec:bc} details the Bohm sheath boundary conditions we use in our simulations.
 
 \subsection{Ion full-F model}
\label{subsec:iongkmodel}
 
Focusing on the electrostatic and neglecting FLR effects with constant and straight magnetic field lines, the ion one-form $\Gamma_i$ \cite{Frei2020} expressed in the gyrocenter coordinates $\bm Z = (\bm R_{g}, \mu, v_\parallel, \theta) $, where $\bm R_g $ is the gyrocenter position, $\mu = m_i v_\perp^2/(2 B)$ is the magnetic moment and $v_\parallel = \bm b \cdot \bm v$ is the velocity parallel to the magnetic field with $\bm b = \bm B / B$, reduces to
 
\begin{align} \label{eq:Gammai}
\Gamma_i(\bm R_g, \mu, v_\parallel,t)= q_i \bm A_i^* \cdot \dot \R_g - \frac{\mu B}{\Omega_i} \dot \theta - m_i v_\parallel^2 / 2 - q_i \Phi_i,
\end{align}
 \\
 with $q_i \Phi_i  =q_i \phi + \mu B$ and $q_i \bm A_i^* = q_i \A +
 m_i v_\parallel \bm b$. We remark that, in \eqref{eq:Gammai}, the electrostatic potential $\phi$ is evaluated at the gyrocenter position, i.e. $\phi = \phi(\bm R_g)$, such that ion FLR effects are neglected. From \eqref{eq:Gammai}, we deduce the ion equations of motion, 

\begin{subequations} \label{eq:ionequationofmotion}
 \begin{align} 
  \dot{\bm R_g} &  =  \bm b v_\parallel +  \frac{\E \times \bm b}{B}  , \label{eq:dotR} \\
\dot v_\parallel & = \frac{q_i}{m_i   } \bm b \cdot \bm E \label{eq:dotvpar},
 \end{align}
 \end{subequations}
 \\
and $\dot \mu =0$, with $v_\parallel = \dot{\R_g} \cdot \bm b$ and $\E = -\grad \phi$ the electric field. \eqref{eq:dotR} describes the parallel streaming along the magnetic field lines and the perpendicular drift due to the $\E \times \B$ velocity, while \eqref{eq:dotvpar} represents the acceleration in the parallel direction associated with the electric field $\bm E$. Using the equations of motion given in \eqref{eq:ionequationofmotion}, the evolution equation of the full-F (gyrophase-independent) ion distribution function, $ F_i = F_i (\bm R_g, \mu, v_\parallel, t)$, in the electrostatic limit of the GK ion Boltzmann equation \cite{Brizard2007} is given by
  
 \begin{align} \label{eq:GKBion}
     \frac{\partial}{\partial t} \left( \J_i F_i \right)   + \grad \cdot \left( \J_i \dot \R_g F_i \right) + \frac{\partial}{\partial v_\parallel} \left(  \J_i F_i \dot v_\parallel \right) = \J_i \C_{i} + \J_i S_i, 
 \end{align}
 \\
 where $\J_i = B / m_i $ is the gyrocenter phase-space Jacobian, which is a constant in the case of linear devices. On the right-hand side of \eqref{eq:GKBion}, $S_i = S_i(\bm R_g, \mu , v_\parallel) = S_{\N} + S_{\mathcal{E}} $ models the particle ($S_N$) and energy ($S_E$) sources and are defined by \cite{Goerler2011,Sarazin2010}

\begin{subequations}\label{eq:sources} 
     \begin{align} 
 S_{\N} &  = \mathcal{A}_\N F_{Mi}, \\
 S_{\mathcal{E}} &  = \mathcal{A}_\mathcal{E} \left( s_{\| i}^2 + x_i - \frac{3}{2}\right) F_{Mi}, 
     \end{align}
\end{subequations}
     \\
  respectively. We note that the sources, given in \eqref{eq:sources}, neglect FLR and polarization effects associated with the transformation from particle to gyrocenter coordinate \cite{Wiesenberger2022}. In \eqref{eq:sources}, the perpendicular normalized velocity-space coordinate is defined by $x_i = \mu B
 /T_{i0}$, while we use $s_{\parallel i}= (v_\parallel  - U_{\parallel i})/ v_{Ti}$ (with $v_{Ti}^2 = 2 T_{i0} /m_i$ and $U_{\parallel i} = \bm b \cdot \bm u_i = \int d \bm v F_i v_\parallel / N_i$ the ion parallel fluid velocity) for the normalized parallel velocity-space coordinate. Here, $T_{i0}$ is the reference ion temperature, which is assumed constant in time and space. We remark that in previous fluid investigations of LAPD, the low ion temperature assumption ($T_i \ll T_e$) is used and the ion energy source is neglected.

 The functions $\mathcal{A}_\N =  \mathcal{A}_\N(x,y)$ and $ \mathcal{A}_\mathcal{E} = \mathcal{A}_\mathcal{E} (x,y)$ in \eqref{eq:sources} are chosen to describe the spatial localization of the sources resulting from ionization processes due to fast electrons and ions \cite{Tripathi2011}. Neglecting the presence of localized sources (not uniform in $z$) near the end plates, we assume that these sources are a uniform in $z$ and radially localized with a top-hat-like shape. For instance, $\mathcal{A}_\N (x,y)$ is given by \cite{Rogers2010}
     
     \begin{align} \label{eq:ANxy}
     \mathcal{A}_\N (x,y)  = \mathcal{A}_{\N0}  0.5  \left[ 1 - \tanh\left( \frac{r -r_s}{L_s}\right)  \right] + \mathcal{A}_{\N \infty},
     \end{align}
     \\
where $r = \sqrt{x^2 + y^2}$ is the perpendicular distance from the center of the device ($r =0$), $r_s$ is the radial extent of the plasma source, $L_s > 0 $ is its typical source decay scale length, $\mathcal{A}_{N0}$ is a positive and constant coefficient, which represents the physical particle fuelling rate near the center of the device, while $\mathcal{A}_{\N \infty} $ represents a small ($\mathcal{A}_{\N \infty } \ll \mathcal{A}_{\N0}$) positive and constant particle source away from $r \sim r_s$ added for numerical reasons, in particular, to avoid regions of small or negative plasma density. Similar definitions for $\mathcal{A}_\mathcal{E}(x,y)$, $\mathcal{A}_{\mathcal{E}0}$ and $\mathcal{A}_{\mathcal{E}\infty}$ are used. In \eqref{eq:sources}, we also introduce a shifted Maxwellian distribution function defined by \cite{Brizard1992,Madsen2013}
  
  \begin{align} \label{eq:fmi}
    F_{Mi}  = \frac{N_i}{ \pi^{3/2} v_{Ti}^3} e^{-
s_{\parallel i}^2} e^{- x_i},
\end{align}
\\
which assumes isotropic ion temperatures. We remark that the sources given in \eqref{eq:sources} neglect non-Maxwellian sources, which can be relevant to fast ion injections, and that the functional form of \eqref{eq:sources} has a straightforward projection in the Hermite-Laguerre basis.

Finally, the term $\C_{i}$ in \eqref{eq:GKBion} is a full-F and nonlinear collision operator model describing ion-ion collisions. In particular, we use a long-wavelength (neglecting FLR effects) Dougherty collision operator \cite{Dougherty1964}, given by

\begin{equation}\label{eq:doughertyfokkerplank}
    \C_{i} =   \nu_{i} \frac{\partial}{\partial \bm v} \cdot \left[ \frac{2 T_i}{m_i}  \frac{\partial}{\partial \bm v}{F_i}-  (\bm v - \bm u_i) F_i \right],
\end{equation}
     \\
     where $T_i =  \int d \vi F_i m_i ( \bm v - \bm u_i)^2 /( 3N_i) $ and $\bm u_{i} =  \int d \vi F_i \bm v / N_i$ are the ion temperature and mean fluid velocity, and $\nu_i =  4 \sqrt{\pi} N_i q_i^4 \ln \lambda/(3 m_i^{1/2} T_{i0}^{3/2})$ is the ion-ion collision frequency, which is constant in the present work. The effects of ion-electron collisions are neglected in \eqref{eq:GKBion} since they occur on a time scale larger by, at least, a factor proportional to $\sqrt{m_i / m_e}$ than ion-ion collisions. 

\eqref{eq:GKBion} is equivalent to the ion GK model used in previous GK turbulent simulations of LAPD, implemented in the \verb|Gkeyll| \cite{Shi2017} and in the \verb|GENE| \cite{Pan2018} codes. Both implementations use the same nonlinear Dougherty collision operator for ion-ion collisions (given in \eqref{eq:doughertyfokkerplank}) and neglect ion-electron collisions. The \verb|Gkeyll| code employs a discontinuous-Galerkin approach and \verb|GENE| uses a finite-volume method to discretize the velocity-space coordinates $(v_\parallel, \mu)$, while our work uses the GM approach to simulate the full-F ion distribution function $F_i$. To our knowledge, this is the first time such a moment approach is applied to perform nonlinear full-F turbulent simulations with an arbitrary number of ion GMs.
   
\subsection{Full-F ion GM Hierarchy Equation}
\label{subsec:fullFhierarchy}

Following Ref. \onlinecite{Frei2020}, we perform the GM expansion of the full-F ion distribution
function, $F_i$. More precisely, we expand $F_i$ onto a set of Hermite, $H_p(s_{\parallel i})$, and Laguerre, $L_j(x_i)$, velocity-space polynomials \cite{Gradshteyn}, such that
 
\begin{align} \label{eq:fullFi}
    F_i = \sum_{p=0}^{\infty} \sum_{j=0}^{\infty} \N^{pj} \frac{H_p(s_{\parallel i}) L_j(x_i)}{\sqrt{2^p p!}}  \frac{F_{Mi}}{N_i},
\end{align}
\\
where $\N^{pj}$ are the ion GMs, evaluated by using the Hermite and Laguerre orthogonality relations \cite{Gradshteyn}

\begin{align} \label{eq:Nipjdef}
  \N^{pj}   = 2 \pi \int_{- \infty}^\infty d v_\parallel \int_{0}^\infty d \mu \frac{B}{m_i } F_i \frac{H_p(s_{\| i}) L_j(x_i)}{\sqrt{2^p p!}}.
\end{align}
\\
By introducing the GM projector \cite{Madsen2013,Jorge2017,Mandell2018,Frei2020}

\begin{align} \label{eq:momentprojector}
    \projpj{pj}{ \chi}  =  2 \pi \int_{- \infty}^\infty d v_\parallel \int_{0}^\infty d \mu \chi \frac{B}{m_i} F_i \frac{H_p(s_{\| i}) L_j(x_i)}{\sqrt{2^p p!}},
\end{align}
\\
 with $\chi = \chi(\bm R_g, \mu, v_\parallel, t)$ being an arbitrary gyrocenter phase-space function, we find $ \mathcal{N}^{pj} = \projpj{pj}{1}$ from \eqref{eq:momentprojector}. We remark that, in \eqref{eq:fullFi}, the shift of the parallel velocity coordinate $s_{\| i}$, appearing in $F_{Mi}$ defined in \eqref{eq:fmi} and in the argument of the Hermite polynomial $H_p$, is necessary to ensure good convergence property of the GM approach with respect to the number of GMs in \eqref{eq:fullFi}, in particular in the presence of sonic ion flows (see \secref{sec:vsp}). These flows appear at the sheath entrance where ions are accelerated to the ion sound speed (see \secref{subsec:bc}). Additionally, we note that $F_{Mi}$, defined in \eqref{eq:fmi}, is assumed to have the same parallel and perpendicular temperature, $T_{\parallel i} = T_{\perp i} = T_{i0}$. The assumption of an isotropic Maxwellian distribution function in \eqref{eq:fullFi} is justified by the large ion-ion collision frequency typically found in a linear plasma device (where $T_i \lesssim 1$ eV) compared to the boundary region in fusion devices (where $T_i \gtrsim 10$ eV). The absence of strong external energy sources driving temperature anisotropy in LAPD experiments supports this assumption (see \eqref{eq:sources}). 

The lowest-order GMs can be related to fluid ion gyrocenter quantities,
such as the ion gyrocenter density $N_i $, the ion
parallel velocity $U_{\parallel i}$, and the ion parallel and perpendicular pressure and temperature $P_{\parallel i} = T_{\parallel i} N_i$ and $P_{\perp i} = T_{\perp i} N_i$, respectively. Indeed, using \eqref{eq:Nipjdef},
we derive that 

\begin{subequations}  \label{eq:fluid2GMs}  
    \begin{align}
    N_i   & = \N^{00} , \\
\N^{10} & =0 , \label{eq:N10} \\
P_{\parallel i} = N_i    T_{\parallel i}  & =  2 \pi \int_{- \infty}^\infty d v_\parallel \int_{0}^\infty d \mu  B F_i(v_\parallel - U_{\parallel i})^2 \nonumber \\
& = T_{i0} \left(  \sqrt{2} \N^{20}    + N_i \right), \label{eq:Tpara} \\
P_{\perp i} = N_i    T_{\perp i}  & =  2 \pi \int_{- \infty}^\infty d v_\parallel \int_{0}^\infty d \mu  \frac{B}{m_i} \mu B F_i \nonumber \\ 
& = T_{i0} ( N_i - \N^{01}),\label{eq:Tperp}
\end{align}
\end{subequations}
\\
with the total ion temperature defined by $T_i = \int d \vi F_i m_i ( \bm v - \bm b U_{\parallel i})^2 /( 3N_i) = (T_{\parallel i} + 2
T_{\perp i})/3 = T_{i0} ( \sqrt{2} \N^{20} + 3 N_i - 2 \N^{01})/( 3N_i )$. We remark that \eqref{eq:N10} is a direct consequence of our choice of using a shifted parallel velocity-space coordinate $s_{\| i}$ in \eqref{eq:fullFi}. In Appendix \ref{appendixA}, we use the definitions of \eqref{eq:fluid2GMs} to derive the Braginskii ion temperature equation. We note that the ion gyrocenter fluid quantities given in \eqref{eq:fluid2GMs} are equivalent to the particle ones since FLR and polarization effects \cite{Jorge2017,Jorge2019,Wiesenberger2022}, associated with the transformation from particle to gyrocenter coordinates, are neglected in our model.

We now derive the full-F GM hierarchy equation describing the evolution of an arbitrary number of GMs, $\N^{pj}$.  This is obtained by projecting the ion full-F equation given in \eqref{eq:GKBion} onto the Hermite-Laguerre basis. In addition, we normalize time $t$ to $ R / c_{s0}$ (with $c_{s0} = \sqrt{T_{e0} / m_i}$ the ion sound speed evaluated at the reference constant electron temperature $T_{e0}$ and $R$ the radial extension of the plasma chamber in the direction perpendicular to $\bm B$), the potential $\phi$ to $ T_{e0} / e$, the parallel and perpendicular spatial scales to $R$ and $\rho_{s0} = c_{s0} / \Omega_i$, respectively. We also normalize the ion and electron densities, $N_i$ and $N_e$, to the constant reference density $N_{0}$, the parallel electron velocity $U_{\parallel e}$ to $c_{s0}$, and the electron temperature, $T_e$, to $T_{e0}$. In addition, we assume $q_i =  + e$, considering a hydrogen plasma. Hence, we derive the normalized ion GM hierarchy equation, which describes the evolution of the GMs $\N^{pj}$, i.e. 

\begin{align} \label{eq:ionhierarchy}
    & \frac{\partial}{\partial t} \N^{pj}  + \sqrt{\frac{p}{\tau_i}} \N^{p-1j} \frac{\partial}{\partial t} U_{\| i}  + \grad \cdot \projpj{pj}{ \dot \R_g}   \nonumber \\ 
  &  + \sqrt{\frac{p}{\tau_i }} \projpj{p-1 j}{\dot{\R_g}}\cdot\grad U_{\| i} - \sqrt{\frac{p}{\tau_i}}  \projpj{p-1j}{ \dot v_\parallel} =  \C^{pj}_i + S_\N^{pj} + S_\mathcal{E}^{pj},
\end{align}
\\
where the GM projections are given by

\begin{subequations} \label{eq:projections}
\begin{align}
  \grad \cdot\projpj{pj}{ \dot{\R_g}} & =\sqrt{\tau_i}  \partial_z
\left(\sqrt{p+1}\N^{p+1 j}+\sqrt{p}\N^{p-1 j}\right) \nonumber \\ 
  &  +  \partial_z  \left( U_{\| i}  \N^{ pj} \right) +\frac{1}{\rho_*} \poissonbracket{\phi}{\N^{ p j}} \label{eq:dotRpj}, 
    \end{align}
    \begin{align}
&     \sqrt{\frac{p}{\tau_i}}\projpj{p-1j}{\dot{\bm{R}_g}} \cdot \grad U_{\| i } =  \left(  p\N^{pj} + \sqrt{p(p-1)} \N^{p-2j}  \right. \nonumber \\ 
     & \left. + \sqrt{\frac{p}{\tau_i}} U_{\| i} \N^{p-1j}\right)  \partial_z U_{\| i}  +   \sqrt{\frac{p}{\tau_i}} \frac{1}{\rho_*} \N^{ p-1 j} \poissonbracket{ \phi}{U_{\| i}}  ,  \end{align}
        \begin{align}
      \sqrt{\frac{p}{\tau_i}}  \projpj{p-1j}{ \dot v_\parallel}  & =   -  \N^{p-1j} \sqrt{\frac{p}{\tau_i}} \partial_z \phi,
    \end{align}
    \end{subequations}
\\
with $\rho^* = \rho_{s0} / R$ and $\tau_i = T_{i0} / T_{e0}$. In \eqref{eq:projections}, we introduce the Poisson bracket operator that is $\poissonbracket{f}{g} = \partial_x f \partial_y g - \partial_y f \partial_x g$. The GM expansions of the particle and energy sources, $S_N^{pj}$ and $S_{\mathcal{E}}^{pj}$, are given by

\begin{subequations} \label{eq:sourcesnpj}
\begin{align} \label{eq:densitysource}
S_\N^{pj}  =   \mathcal{A}_\N \delta_p^0  \delta_j^0 = S_\N,
\end{align}
\begin{align}
S_\mathcal{E}^{pj}  = \mathcal{A}_\mathcal{E} \left(   \frac{\delta_p^2  \delta_j^0}{\sqrt{2}}  -  \delta_p^0  \delta_j^1 \right),  \label{eq:energysource}
\end{align}
\end{subequations}
\\
respectively. The contribution of \eqref{eq:densitysource} to the ion density equation obtained with $(p,j) =(0,0)$ represents particle sources, while the non-vanishing terms with $(p,j)=(2,0)$ and $(0,1)$ in \eqref{eq:energysource} are associated with parallel and perpendicular energy sources.

Finally, we express the nonlinear Dougherty collision operator in terms of GMs. We first express \eqref{eq:doughertyfokkerplank} in terms of the velocity-space coordinates $(s_{\| i}, x_i)$ and project it onto the Hermite-Laguerre basis. This yields

\begin{align}\label{eq:moments:nonlinear_dougherty}
   \C^{pj}_i & = \nu_{i}\left[ -(p+2j) \N^{pj} + \left(T_i -1 \right) \right. \nonumber \\
  & \left. \times \left(\sqrt{p(p-1)} \N^{p-2j} - 2j   \N^{pj-1} \right) \right],
\end{align}
\\
where $T_i$ is expressed in terms of the GMs using \eqref{eq:fluid2GMs}. The nonlinear Dougherty collision operator conserves particles ($\C_i^{00} =0$), momentum  ($\C_i^{10} =0$) and energy ($ \C_i^{20} = \sqrt{2} \C_i^{01}$). While simpler in form compared to the GM expansion of the nonlinear Fokker-Planck Landau collision operator \cite{Jorge2019}, the Dougherty collision operator constitutes an initial step to incorporate advanced collisional effects in the nonlinear and full-F ion GM hierarchy equation. The numerical implementation of the nonlinear Fokker-Planck Landau collision operator \cite{Jorge2019} will be considered in future work.

To obtain the time evolution of the GMs $\mathcal{N}^{pj}$, it is necessary to derive an explicit expression for the time derivative of the ion parallel velocity, $\partial_t U_{\| i}$ which appears in \eqref{eq:ionhierarchy} and resulting from the use of the shifted parallel velocity-space coordinate $s_{\| i}$. By setting $(p,j) = (1,0)$ in \eqref{eq:ionhierarchy} and using the fact that $\N^{10}$ vanishes exactly (see \eqref{eq:N10}), we derive the desired expression for $\partial_t U_{\| i}$ given by

   \begin{align} \label{eq:dtui}
 &       N_i  \partial_t  U_{\| i}  + \frac{N_i}{\rho_*  } \poissonbracket{ \phi}{U_{\| i}}  +\tau_i  \partial_z P_{\parallel i}+    N_i  U_{\| i} \partial_z  U_{\| i} +  N_i \partial_z \phi =0,
\end{align}
\\
where the parallel ion pressure $P_{\| i}$ is expressed in terms of GMs according to \eqref{eq:Tpara}. 

We note that the full-F GM hierarchy equation, given in \eqref{eq:ionhierarchy}, can also be derived from the electromagnetic full-F GM hierarchy equation described in Ref. \onlinecite{Frei2020}. This is achieved by considering the electrostatic limit, neglecting FLR effects, and assuming isotropic and constant ion temperatures in $F_{Mi}$. We remark that, if $F_{Mi}$ is normalized with the local parallel and perpendicular ion temperatures ($T_{\parallel i}$ and $T_{\perp i}$ in \eqref{eq:Tpara} and \eqref{eq:Tperp}, respectively) as in Ref. \onlinecite{Frei2020}, additional terms proportional to their time derivatives and gradients appear in \eqref{eq:ionhierarchy}. These terms might be important in the case of large temperature variations such as the ones in the boundary, but can be neglected in the case of LAPD.

 Notably, the GMs with different $p$ are coupled in \eqref{eq:ionhierarchy} due to the parallel streaming terms, associated with the ion Landau damping. On the other hand, the GMs with different $j$ are only coupled through the collision operator (see \eqref{eq:moments:nonlinear_dougherty}) as our model neglects FLR effects yielding additional coupling in $j$ \cite{Frei2022}. As a result, a few Laguerre GMs are expected to be sufficient in our nonlinear turbulent simulations. 

To carry out the numerical turbulent simulations presented here, a simple closure by truncation is applied to the GM hierarchy equation. More precisely, we set $\N^{pj} =0$ for all $(p,j) > (P,J)$ with $0 \leq P,J < \infty$. The full-F GM hierarchy equation enables us to perform turbulent simulations of LAPD using an arbitrary number of GMs. Different values of $(P,J)$ are considered in \secref{sec:turbulentsimulations} where we demonstrate that the closure by truncation is sufficient to perform full-F turbulent simulations in our case.

\subsection{Cold-ion reduced model}
\label{subsec:coldion}

We consider here the cold-ion limit of the full-F GM hierarchy and derive a simplified model, similar to the one used in previous turbulent investigations of linear devices based on fluid models (see, e.g., Refs. \onlinecite{Rogers2010,Popovich2010,Fisher2015}) where the effects of finite ion temperature $T_i$ are neglected.

In the cold ion limit ($T_i =0$), only the GM $\N_i^{00}$ and $U_{\| i }$, associated with the ion gyrocenter density and the parallel ion velocity, need to be evolved and the contribution from the parallel ion pressure $P_{\parallel i }$ in \eqref{eq:dtui} can be neglected. As a consequence, the ion GM hierarchy equation given in \eqref{eq:ionhierarchy} reduces to the ion gyrocenter continuity equation for $N_i$ and to the ion parallel momentum equation for $U_{\| i}$ \cite{Wiesenberger2020,Wiesenberger2022,Galassi2022}, i.e. 

\begin{subequations} \label{eq:coldion}
\begin{align} 
&        \frac{\partial}{\partial t} N_i  + \frac{1}{\rho^*} \poissonbracket{ \phi }{ N_i} +   \partial_z    \left( U_{\| i}  N_i \right)   =  S_{\N}, \label{eq:nicoldion} \\
&        \partial_t  U_{\| i}     + \frac{1}{\rho_*} \poissonbracket{ \phi }{U_{\| i}}   + U_{\| i} \partial_z U_{\| i}   + \partial_z \phi =0, \label{eq:uicoldion}
\end{align}
\end{subequations}
\\
respectively. We remark that the particle and momentum conservation of the collision operator is used in deriving \eqref{eq:coldion}.

\subsection{Electron fluid model}
\label{subsec:electronbraginskii}

We use the Braginskii model to evolve the electron dynamics, avoiding the evolution of their distribution function, in contrast to Refs. \onlinecite{Shi2017,Pan2018}. The fluid approach for the electrons is justified when the electron collision frequency is much larger than the ion collision frequency and electron FLR effects are negligible for modes developing at $k_\perp \rho_s \sim 1$, which is the case of LAPD experiments. In the absence of electron FLR effects, the electron particle and gyrocenter position coinicide such that no distinction is made between the electron particle and gyrocenter fluid quantities. Hence, the time evolution of the electron density $n_e$, electron parallel velocity $U_{\parallel e}$, and temperature $T_e$ is determined by the continuity equation, the generalized Ohm's law, and the temperature equation, respectively. These equations are given by 

\begin{subequations} \label{eq:electronbraginskii}
\begin{align} \label{eq:nebraginskii}
\partial_t  n_e  & + \frac{1}{\rho^*} \poissonbracket{ \phi}{  n_e }  + \partial_z \left( U_{\parallel e} n_e \right) =  S_{\N} , \\
        \partial_t U_{\parallel e} & +  \frac{1}{\rho^*} \poissonbracket{ \phi}{  U_{\| e} }   + U_{\parallel e} \partial_z U_{\parallel e} =  \frac{m_i}{m_e} \left[  \nu_\parallel J_\parallel + \partial_z \phi \right.  \nonumber \\
&         \left. - \frac{T_e}{n_e} \partial_z  n_e  - 1.71 \partial_z T_e \right] , \label{eq:uebraginskii}\\
\partial_t  T_e & + \frac{1}{\rho^*}   \poissonbracket{ \phi}{ T_e }  + U_{\parallel e} \partial_z T_e = \frac{2}{3}T_e \left( \frac{0.71}{N_e} \partial_z J_{\parallel} - \partial_z U_{\parallel e} \right) \nonumber \\ 
& + \partial_z \left(   \chi_{\parallel e } \partial_z T_e \right) + S_{T_e} , \label{eq:tebraginskii}
\end{align}
\end{subequations}
\\
where the normalized parallel electrical resistivity and electron thermal conductivity are given by $\nu_\parallel  = \nu_0 / T_e^{3/2}$ and $\chi_{\parallel e} = 1.075 T_e^{5/2} / \nu_0$, respectively. Here, $\nu_0  = 4 \sqrt{2 \pi} e^4 n_{e0}   R \sqrt{m_e} \ln \lambda /[  3  c_{s0}  m_i T_{e0}^{3/2} 1.96  ]$ is the normalized electron collisionality. On the right-hand side of \eqref{eq:nebraginskii} and \eqref{eq:tebraginskii}, $S_{N}$ and $S_{T_e}$ are the normalized density and temperature sources. In \eqref{eq:electronbraginskii}, the parallel electrical current is $J_{\parallel} = n_e ( U_{\| i}   -  U_{\| e} )$.

\subsection{Vorticity equation}
\label{subsec:vorticityequation}

We now obtain the vorticity equation that governs the evolution of the electrostatic potential $\phi$. This equation imposes the charge conservation constraint to the time evolution of the plasma densities and electrical currents.

To derive the vorticity equation, we consider the quasineutrality condition in the long-wavelength limit, given by \cite{Brizard2007,Frei2020} 

\begin{align} \label{eq:quasineutrality}
 - e n_e + q_i N_i  = -  \grad \cdot \left(  \frac{q_i^2 N_i}{m_i \Omega_i^2} \grad_\perp \phi\right).
\end{align}
\\
We remark that \eqref{eq:quasineutrality} neglects FLR effects (associated with the gyro-average) but retains the polarization effect in its right-hand side. Including high-order FLR contributions to the gyro-average and polarization terms, which are important for predicting of small-scale fluctuations \cite{Held2022}, requires the development of FLR closures at arbitrary wavelength (see, e.g., Ref. \onlinecite{Held2020}) using an arbitrary number of GMs, a task left to future work. We also notice that \eqref{eq:quasineutrality} is equivalent to the quasineutrality condition used in previous GK turbulent simulations of LAPD \cite{Shi2017,Pan2018} if the Boussinesq approximation is used, i.e. if $N_i$ is approximated by $N_0$ on the right-hand side of \eqref{eq:quasineutrality}. The Boussinesq approximation is widely used in fluid codes \cite{Ross2019,Giacomin2022} and is also considered below to derive the vorticity equation.

While \eqref{eq:quasineutrality} can be solved to obtain $\phi$ given the electron and ion densities, $n_e$ and $N_i$ respectively, we use a vorticity equation instead which is often considered in turbulent fluid codes \cite{Rogers2010, Popovich2010,Fisher2015,Ross2019}. The vorticity equation is derived by taking the time derivative of the quasineutrality equation given in \eqref{eq:quasineutrality} and by using the electron and ion continuity equations, given in Eqs. (\ref{eq:nicoldion}) and (\ref{eq:nebraginskii}), respectively. Furthermore, we use the Boussinesq approximation, such that

\begin{align}
\grad \cdot \left(  \frac{q_i^2 N_i}{m_i \Omega_i^2} \grad_\perp \phi\right) \simeq \frac{q_i^2 N_i}{m_i \Omega_i^2}  \Omega,
\end{align}
\\
where we introduce the vorticity variable $\Omega = \grad^2_\perp \phi$. The vorticity equation is then 

\begin{align} \label{eq:vorticityboussinesq2}
     & -  \partial_t   \Omega 
    -  \frac{1}{\rho^*}  \poissonbracket{\phi}{ \Omega }      - \partial_z(  U_{\parallel i} \Omega )  +  \frac{1}{N_i}\partial_z  J_\parallel  = 0,
\end{align}
\\
We note that the effects of the Boussinesq approximation on plasma turbulence are the subject of previous studies \cite{Yu2006,Bodi2011,Paruta2018,Ross2019,Held2022}. While it might not be justified in LAPD when steep density gradients are present, it allows us to reduce the computational cost of our simulations when inverting the two-dimensional Laplacian to obtain $\phi$ from the vorticity variable $\Omega$. We use the vorticity equation given in \eqref{eq:vorticityboussinesq2} to evolve $\Omega$ when considering the full-F ion GM hierarchy equation and the cold ion models, given in Eqs. (\ref{eq:ionhierarchy}) and (\ref{eq:coldion}) respectively, coupled to the fluid electron model in \eqref{eq:electronbraginskii}.

\subsection{Two-fluid Braginskii fluid model}
\label{subsec:braginskii}

We finally introduce the two-fluid Braginskii fluid model \cite{Zeiler1997}, valid in the high-collisional regime, for comparison with the full-F ion GM hierarchy equation and the cold-ion model. 

We first note that, in the two-fluid Braginskii model, quasineutrality is assumed (neglecting the polarization term in \eqref{eq:quasineutrality}) and, as a consequence, the electron particle density is used as an independent variable, such that $n_e \simeq N_i$. On the other hand, ion polarization effects are retained by the presence of the polarization drift (neglected in \eqref{eq:dotR}) in the ion particle continuity equation, from which the vorticity equation (given below) is derived for the electrostatic potential. Furthermore, since FLR effects associated with the difference between particle and gyrocenter positions are neglected in the two-fluid Braginskii fluid model, the ion gyrocenter fluid quantities, such as ion parallel velocity $U_{\parallel i}$ and ion temperature $T_i$, are assumed to be equal to particle fluid quantities.

In addition to the fluid electron fluid equations for $n_e$, $U_{\| e}$ and $T_e$ already described in \secref{subsec:electronbraginskii}, the two-fluid Braginskii equations prescribe a parallel ion momentum equation to evolve $U_{\parallel i}$, an ion temperature equation to evolve $T_i$, and vorticity equations for $\Omega$. These equations are given by

\begin{subequations} \label{eq:ionbraginskii}
\begin{align}
\partial_t  U_{\parallel i}  + \frac{1}{\rho^*} \poissonbracket{ \phi}{  U_{\parallel i} } + U_{\parallel i} \partial_z  U_{\parallel i}  & = -  \partial_z T_e -   \tau_i   \partial_z   T_i  \nonumber \\ 
& - (T_e + \tau_i T_i)  \frac{\partial_z  n_e}{n_e} ,  \label{eq:uibraginskii}
\end{align}
\begin{align}
\partial_t  T_i  +  \frac{1}{\rho^*} \poissonbracket{ \phi}{  T_i }    + U_{\parallel i} \partial_z T_i  & = +\frac{2}{3} T_i \left[\left(U_{\parallel i}- U_{\parallel e}\right) \frac{\partial_z  n_e}{n_e}- \partial_z U_{\parallel e}\right]    \nonumber \\ 
&   + \partial_z \left( \chi_{\parallel i} \partial_z T_i \right) +  \frac{\mathcal{A}_\mathcal{E}}{n_e}  +(1 - T_i) \frac{\mathcal{A}_\N}{n_e} \label{eq:tibraginskii},
\end{align}
\begin{align} \label{eq:vorticitybraginskii}
\partial_t \Omega  +  \tau_i  \partial_t \grad_\perp^2 T_i &= \frac{1}{n_e}\partial_z J_{\parallel} -  \frac{1}{\rho^*}  \poissonbracket{ \phi}{  \Omega + \tau_i \grad_\perp^2 T_i } \nonumber \\ 
  &    -U_{\parallel  i} \partial_z \left( \Omega + \tau_i \grad_\perp^2 T_i \right).
\end{align}
\end{subequations}
\\
respectively. In \eqref{eq:tibraginskii}, $\chi_{\parallel i} = 1.32 \sqrt{m_e / m_i} (\tau _i T_i)^{5/2}   / \nu_0$ is the normalized parallel ion thermal conductivity. In \eqref{eq:tibraginskii}, the two last terms are the ion temperature sources associated with the energy source $S_\mathcal{E}$ (see \eqref{eq:energysource}), which appears on the right-hand side of \eqref{eq:GKBion}. In Appendix \ref{appendixA}, we demonstrate analytically how these terms are determined by deriving the ion temperature equation, \eqref{eq:tibraginskii}, from the ion GM hierarchy equation.

In contrast to the cold-ion model given in \eqref{eq:coldion}, the parallel electric field $\partial_z \phi$, appearing in the parallel ion momentum equation, \eqref{eq:uicoldion}, is approximated in \eqref{eq:uibraginskii} by the electron parallel pressure gradient, such that $\partial_z \phi \simeq \partial_z P_e$ with $P_e = n_e T_e$ (see \eqref{eq:uebraginskii}). We also note that the terms proportional to the Laplacian of the ion temperature, i.e. $\tau_i \grad_\perp^2 T_i$, present in the vorticity equation, \eqref{eq:vorticitybraginskii}, are absent in \eqref{eq:vorticityboussinesq2}, which is derived from the GM hierarchy equation. Indeed, these terms are associated with the long-wavelength FLR effect correction, which are neglected in \eqref{eq:vorticityboussinesq2} as FLR effects are omitted. While the ion temperature in LAPD experiments is generally lower than the electron temperature, it has been shown that FLR effects can be important in this regime. Further investigations are hence required to investigate the effects of finite ion temperature and related FLR effects on small scale fluctuations in LAPD.

\subsection{Boundary conditions}
\label{subsec:bc}

Boundary conditions are required for the ion GMs, $\N^{pj}$, the electron fluid quantities, $N_e$, $U_{\parallel e}$, $T_e$, and the potential $\phi$ in the perpendicular $(x,y)$ plane at $x = \pm L_x / 2$ and $y = \pm L_y / 2$ and at the end plates located in the $z$ direction at $z = \pm L_z /2$, where a sheath forms due to the plasma-wall interaction. 

At $x = \pm L_x / 2$ and $y = \pm L_y / 2$, homogenous Neumann boundary conditions are used for all quantities. These \textit{ad-hoc} boundary conditions have a negligible effect on plasma turbulence near the center of the device as they are imposed at a distance sufficiently large from the center of the device. On the other hand, the boundary conditions in the $z$ direction have an important impact since the formation of a Debye sheath is observed when the magnetic field lines intercept the end plates that control the plasma losses \cite{Stangeby1995}. Since the sheath region cannot be modeled by the field equations derived in \secref{subsec:vorticityequation} (the GK formalism is violated in this region), the sheath is modeled in our simulations by a set of appropriate boundary conditions imposed at the sheath entrance.

In previous GK simulations of LAPD \cite{Shi2017,Pan2018}, a conducting wall is considered. Accordingly, the fraction of electrons that cross the sheath and are lost being absorbed by the walls is determined by the value of the potential at the sheath entrance. This fraction is imposed by evaluating the cutoff velocity of the electron distribution function numerically. Leveraging the GM approach, we use the standard fluid Bohm boundary conditions \cite{Stangeby1995} which sets the value of the parallel electron and ion velocities, $U_{\parallel e}$ and $U_{\parallel i}$, at the sheath entrance. Therefore, we assume that \cite{Loizu2012,Mosetto2015}

\begin{subequations} \label{eq:vparbc}
\begin{align} 
U_{\parallel e}(x,y,z =  \pm L_z / 2) &= \pm \sqrt{T_{e,s}} e^{   \Lambda  - \phi_s / T_{e,s}}, \\
U_{\parallel i}(x,y,z  = \pm L_z / 2)  &=  \pm c_s & \nonumber \\
& = \pm \sqrt{T_{e,s}} \sqrt{1 + \tau_i T_{i,s} / T_{e,s}}, \label{eq:vparibc}
\end{align}
\end{subequations}
\\
with $\Lambda = \log m_i/(2m_e) \simeq 3$ for hydrogen plasmas. In \eqref{eq:vparbc}, $T_{e,s}$ and $T_{i,s}$ are the electron and ion temperatures evaluated at the sheath entrance, i.e. $T_{e,s} = T_e(x,y, z = \pm L_z / 2)$ and $T_{i,s} = T_i(x,y, z = \pm L_z / 2)$, and, similarly $\phi_s = \phi(x,y, z =\pm L_z / 2)$. We notice that the boundary conditions in \eqref{eq:vparbc} reduce to the ones used in Ref. \onlinecite{Rogers2010} when $T_i \ll T_e$ and correspond to the ones used in SOL turbulent simulations using the drift-reduced Braginskii model \cite{Giacomin2022}. 
For the remaining quantities, we assume, for simplicity, that the gradients of electron density, $n_e$, electron temperature, $T_e$, ion GMs, $\N^{pj}$, and electrostatic potentials, $\phi$, vanish along the direction of the magnetic field at the sheath entrance, i.e. homogenous Neumann boundary conditions are imposed at $z = \pm L_z / 2$. While the homogenous Neumann boundary conditions considered here are sufficient to ensure the numerical stability of the present simulations, further investigations are needed to develop first-principles sheath boundary conditions for the GM approach. In particular, the analytical procedure outlined in, e.g., Refs. \onlinecite{Loizu2012,Mosetto2015}, can be extended to an arbitrary number of GMs and kinetic sheath boundary conditions can also be
developed \citep{Riemann1981,Baalrud2015,Churchill2017}. Magnetic field lines
intercept the machine wall with a small oblique angle in fusion devices, further complicating the treatment of the sheath boundary conditions \cite{Chodura1982,Geraldini2017}. 

\section{Numerical Implementation}
\label{sec:numericalimplementation}

To solve the full-F ion GM hierarchy given in \eqref{eq:ionhierarchy} coupled with the electron fluid model in \eqref{eq:electronbraginskii} and the vorticity equation in \eqref{eq:vorticityboussinesq2}, we have developed a new three-dimensional full-F simulation code. This code solves the turbulent dynamics for an arbitrary number of ion GMs. It also solves the cold ion model (\secref{subsec:coldion}) and the two-fluid Braginskii model (\secref{subsec:braginskii}) for comparison with the GM results.

To evolve the plasma dynamics, we employ similar numerical algorithms as the two-fluid \verb|GBS| code \cite{Giacomin2022}. More precisely, an explicit fourth-order Runge-Kutta time-stepping scheme is used. The perpendicular and parallel directions are discretized using a uniform cartesian grid in the $(x,y,z)$ coordinates with the $x$, $y$, and $z$ directions discretized using $N_x$, $N_y$ and $N_z$ points uniformly distributed between the intervals $[- L_x /2, + L_x /2]$, $[- L_y /2, + L_y /2]$ and $[- L_z /2, L_z/2]$, respectively. The Poisson bracket operator, $\left[ f,g\right] = \bm b \times \grad f \cdot \grad g = \partial_x f \partial_y g - \partial_y f \partial_x g$, with $\bm b = \bm B / B = \bm e_z$, is evaluated by using a fourth-order Arakawa method \citep{Arakawa1997}. The numerical evaluation of the other spatial operators appearing in the GM hierarchy equation is based on a fourth-order and centered finite difference scheme, resulting in a $5$-points centered stencil \cite{Giacomin2022}. To avoid checkerboard patterns \citep{Paruta2018}, the grid used to evolve the parallel velocities, $U_{\parallel e}$ and $U_{\| i}$, and the GMs $\N^{pj}$ with odd $p$, is staggered to the left along the $z$-direction by $\Delta z /2$ ($\Delta z$ is the grid spacing) with respect to the grid where the other fluid quantities, i.e. $n_e$, $T_e$, $\Omega$ (and thus $\phi$), and the GMs $\N^{pj}$ with even $p$ are evaluated. Fourth-order interpolation techniques are used between two staggered grids \cite{Paruta2018}. To improve the numerical stability of our numerical simulations, parallel and perpendicular numerical diffusions, such as

\begin{align}
D(f) = \eta_\perp \left(\partial_{xx}^2 +  \partial_{yy}^2\right) f + \eta_z \partial_{zz}^2 f,
\end{align}
\\
where $f$ denotes one of the evolved quantities, are added to the right-hand side of all equations. We choose the perpendicular and parallel diffusion coefficients, $\eta_\perp$ and $\eta_z$, to be constant and sufficiently small not to affect significantly the results. The model is implemented in a Fortran code using a MPI domain decomposition in all spatial directions. As a consequence,, the computation cost of our simulations scales approximatively linearly with the number of GMs taken into account.

The initial conditions of the turbulent nonlinear simulations impose equal electron and ion densities and temperatures, such that $ n_e = \N^{00}$ and $T_e = T_i$ with top-hat-like profiles in the perpendicular plane and uniform in $z$, similar to \eqref{eq:ANxy}, such that $ f =   \mathcal{A}_0 [1 - \tanh\left\{ (r - r_s) / (4 L_s) \right\}] / 2 $. We use $\mathcal{A}_0 = 0.4$ for $f = n_e$ and $\mathcal{A}_0 = 0.4$  for $f = T_e$. In addition, we set $\phi = \Lambda T_e$ to avoid unphysical and large electron current into the sheath region. The initial values of $\N^{20}$ and $\N^{01}$, given the initial ion density $\N$ and ion temperature $T_i$ profiles, are obtained by inverting \eqref{eq:fluid2GMs}, which yields $\N^{20} = N_i ( T_i- 1) / \sqrt{2}$ and $\N^{01} = N_i ( 1 - T_i) $, with $T_{\parallel i} = T_{\perp i} = T_i$. Finally, the parallel velocities, $U_{\| i}$ and $U_{\| e}$, are initialized with smooth profiles along $z$, interpolating the values at the end plates fixed according to the boundary conditions given in \eqref{eq:vparbc}. Random noise (with constant amplitude $0.01$) is added to the initial densities, $n_e$ and $N_i$, temperatures $T_i$ (and associated GMs $\N^{01}$, $\N^{02}$), and parallel velocities, $U_{\parallel e}$ and $U_{\parallel i}$, profiles to seed turbulence. Typically, a quasi-steady state is achieved after $100 $ $c_{s0} / R$ time unit (corresponding to $t \sim 4$ ms), where the sources of particle and energy are compensated by the losses at the end plates. 

\section{Full-F Turbulent simulation results}
\label{sec:turbulentsimulations}

In this section, we present the first turbulent and full-F simulations of the GM approach of a linear plasma device, focusing on the parameters of the LAPD experiment. We perform a comparison between the turbulent predictions of the full-F GM approach (see \secref{subsec:fullFhierarchy}), with different numbers of GMs and values of collisionality and compare them with the Braginskii model introduced in \secref{subsec:braginskii}.

Our simulations parameters are similar to those used in Ref. \onlinecite{Rogers2010}, where a helium LAPD plasma is considered. These parameters are sumarized as follows: $n_{e0} = 2 \times 10^{12}$ cm$^{-3}$, $T_{e0} = 6$ eV, $T_{i0} = 3 $ eV ($\tau_i = 0.5$), $\Omega_{i} \sim 960$ kHz, $\rho_{s0} = 1.4$ cm, $c_{s0}= 1.3 \times 10^{6}$ cm s$^{-1}$, $m_i / m_e =  400$ and $\nu_0 = 0.03$. The LAPD vacuum chamber has a radius $R \simeq 0.56$ m (i.e., $R \simeq 40 \rho_{s0}$) and a parallel length of $L_z \simeq 18$ m, such that we use $L_x = L_y = 100 \rho_{s0} $ (or $L_x \sim L_y \sim 1.4 $ m) and $L_z = 36 R$. The reference time is $R / c_{s0}  \sim 43$ $\mu$s. We consider the following parameters for the density and temperature sources $L_s = 1  \rho_{s0}$, $r_s = 20 \rho_{s0}$, $\mathcal{A}_{\N0} =  \mathcal{A}_{T_{e0}}  = 0.04 $ (with $\mathcal{A}_{\N\infty} = \mathcal{A}_{T_e \infty}  =  0.001$) and $\mathcal{A}_{\mathcal{E}0}  = 0.02$ (with $\mathcal{A}_{\mathcal{E}\infty} = \mathcal{A}_{\N\infty}$). We use a numerical resolution of $N_x = N_y = 192$ in the perpendicular plane, which corresponds to a grid spacing of $\sim 0.52 \rho_{s0}$. This resolution is also sufficient to resolve turbulence occurring at perpendicular wavelengths longer than $\rho_{s0}$ in the perpendicular plane, which is dominated by the long-wavelength KH instability \cite{Rogers2010}. On the other hand, a coarser resolution of $N_z = 64$ is used in the parallel direction. Given our numerical resolution, we use $\eta_\perp = 0.05 $ and $\eta_z = 5$ for numerical stability. We remark that further numerical studies are required to fully assess the impact of the numerical diffusion and resolution we use in our simulations, but the values considered here are sufficient for the goals of the present work.

In order to investigate the impact of ion collisions, we conduct a set of nonlinear simulations in the high (HC) and low (LC) ion collisionality regime. For each set, we consider different numbers of GMs $(P,J)$ to investigate the convergence of the GM approach. More precisely, we consider $(P,J) = (2,1)$, $(6,1)$, and $(12,1)$ in the LC regime and $(P,J) = (2,1)$ and $(6,1)$ in the HC regime. We change the ion collisionality by varying the ion collision frequency $\nu_i$ as an independent parameter while keeping all other parameters constant. In the HC regime, the ion collision frequency is computed using the LAPD physical parameters, such that $\nu_i = 1.38 \sqrt{m_i / m_e} \nu_0 /\tau_i^{3/2} \simeq 2.34$. In this regime, the ion mean-free-path, $\lambda_{mpf}$, is considerably shorter than the total length $L_z$, i.e.  $\lambda_{mpf} / L_z \simeq \sqrt{2 \tau_i} R / L_z / \nu_i \ll 1$, and the effects of the collision operator are expected to be important. On the other hand, we set the ion collision frequency to be small in the LC regime, such that $\nu_i \simeq 4 \times 10^{-3}$ yielding $\lambda_{mpf} / L_z \sim  6.9$. In this regime, the effect of the collision operator on the GMs is expected to be negligible. In all cases, no artificial diffusion in velocity-space is added, since the damping of the GMs by the Dougherty collision operator given in \eqref{eq:moments:nonlinear_dougherty} is sufficient to ensure stability of our simulations.

We remark that using $J=1$ is expected to be sufficient to represent the ion distribution function $F_i$ in $x_i$ since kinetic effects such as magnetic drifts \citep{Frei2022} and trapped particles \citep{Frei2023}  (which couple GMs with different $j$) are absent in the LAPD configuration. In addition, neglecting the gyro-average in our ion model allows us to remove the coupling between different $j$ driven by the $x_i$-dependence of the gyro-average operator \citep{Frei2020}. We also note that a number $J=1$ of Laguerre polynomials fully resolves the density and energy sources (see \eqref{eq:sourcesnpj}). As a matter of fact, the only coupling to the $j \geq 2$ GMs is from the pitch-angle scattering contained in the collision operator given in \eqref{eq:moments:nonlinear_dougherty}, which is dominated by collisional damping. Indeed, if collisions are neglected GMs with $j >1$ are completely decoupled from the the GMs with $j \leq 1$

This section is structured as follows. First, \secref{sec:simulationresults} provides an analysis and comparison of simulations based on the full-F GM hierarchy, the cold-ion, and the Braginskii models. Second, the turbulence characteristics are analyzed and compared in more details in \secref{sec:turbulence}. Finally, we investigate the ion distribution function in velocity-space in \secref{sec:vsp} and the GM spectrum in quasi-steady state in \secref{subsec:GMspectrum} as a function of the number of GMs and for the two collisionality regimes.

\subsection{Simulation results}
\label{sec:simulationresults}
This section presents a set of nonlinear and turbulent simulations of the LAPD using the full-F GM hierarchy equation given in \eqref{eq:ionhierarchy}, the cold-ion model in \eqref{eq:coldion}, and the Braginskii model introduced in \eqref{eq:ionbraginskii}. 

\begin{figure}
\centering
\includegraphics[scale =0.82]{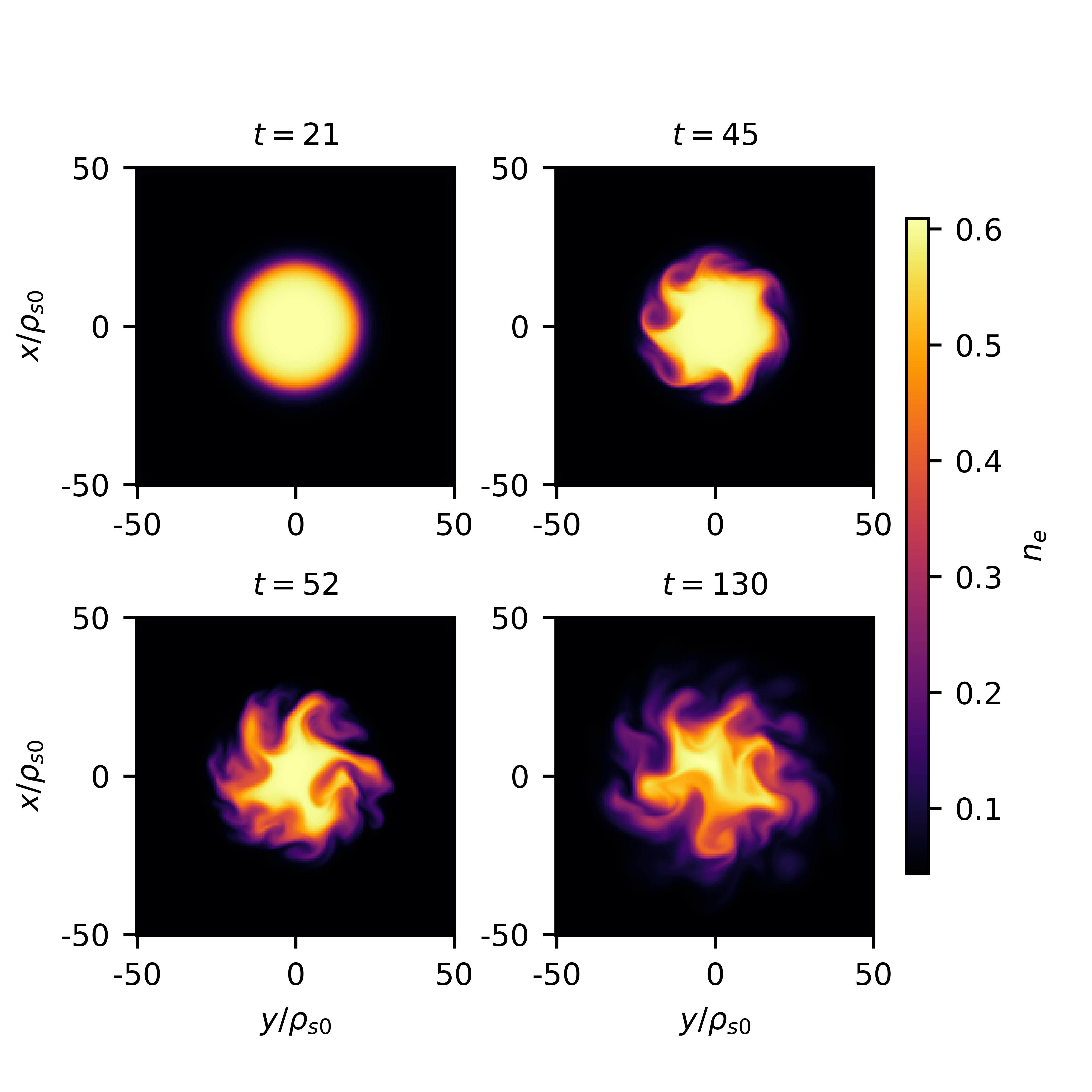}
\caption{Typical snapshots during the time evolution (from top-left to bottom-right) of a LAPD simulation using $(P,J) = (6,1)$ GMs in the HC regime. The normalized electron density $n_e$ (taken at $z=0$) is shown. First, the top-hat-like sources fuel the central region ($r \lesssim r_s$) while a resistive drift-wave instability is triggered as the gradients build up (top left). Second, the KH instability develops (top right), saturates nonlinearly (bottom left), and broadens the profiles to reach a quasi-steady state (bottom right). $n_e$ is normalized, as well as all the quantities shown in the following figures.}
\label{fig:snapshotsne}
\end{figure}

A typical nonlinear evolution of the electron density, $n_e$, obtained by using the GM hierarchy equation with $(P,J) = (6,1)$ GMs in the HC regime is shown in \figref{fig:snapshotsne}. For $t \lesssim 21 R / c_{s0}$, the profiles build up because of the localized particle and energy sources present in the system. The steep density and temperature gradients near $r \sim r_s$ drive an unstable resistive drift-wave (not visible in \figref{fig:snapshotsne} due to their small relative amplitude), with the most unstable mode occurring at $k_\perp \rho_{s0} \sim 0.5$ ($k_\perp$ is the perpendicular wavenumber) with finite parallel wavenumber and rotating in the ion diamagnetic direction. Large poloidal flows, with velocity typically larger than the phase-velocity of the resistive drift waves \cite{Popovich2010}, nonlinearly trigger a KH instability, characterized by a long perpendicular wavelength ($k_\perp \rho_{s0}  \ll 1$) and $k_\parallel \simeq 0$. The KH instability becomes clearly visible around $t \simeq 45 R / c_{s0}$. This instability, which has been shown to dominate the radial transport in LAPD \cite{Rogers2010}, saturates at $t \simeq 53 R / c_{s0}$, transporting the plasma to the $r \gtrsim r_s$ region and yielding the broadening of the initial profiles. The role of the KH-dominated transport in our simulations is confirmed by the strong steepening of the profiles (similar to Ref. \onlinecite{Rogers2010}) when the nonlinear term $\poissonbracket{\phi}{\Omega}$ in \eqref{eq:vorticityboussinesq2} is artificially suppressed. 

\begin{figure}
\centering
\includegraphics[scale =0.62]{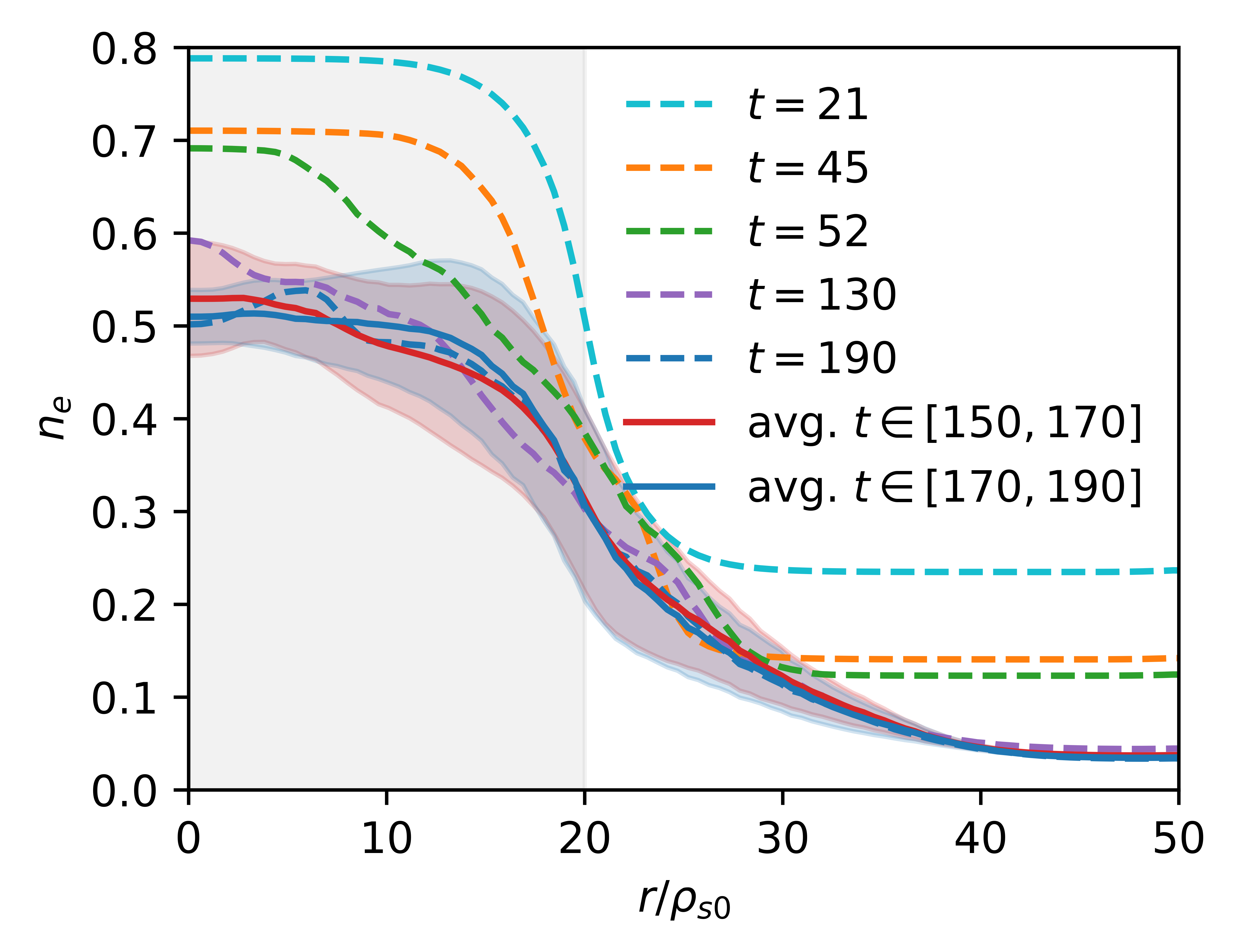}
\caption{Saturation of the radial profile of \(n_e\) (binned by radius) associated with \figref{fig:snapshotsne} at \(z = 0R\). Instantaneous radial profiles are shown by the dashed lines for different times, and time-averaged radial profiles over the time-windows \(t \in [150, 170]\) (red) and \(t \in [170, 190]\) (blue) are also presented for comparison. The shaded colored areas represent the standard deviations of each time-averaged profile. A quasi-steady state is reached for \(t \gtrsim 130\). The time \(t\) is normalized to \(R / c_{s0}\).}
\label{fig:steadysteate}
\end{figure}

After $t \sim 130 R / c_{s0}$ (similarly to previous GK and Braginskii turbulent simulations of LAPD \cite{Popovich2010,Pan2018}), a quasi-steady state is reached. At this point when the sources are compensated by the losses at the end plates. In particular, the quasi-steady state in our simulations can be monitored by examining the radial profiles. This is shown in \figref{fig:steadysteate}, where the initial radial profile of $n_e$ (associated with \figref{fig:snapshotsne}) reaches a quasi-steady state after $t \sim 130 R / c_{s0}$. To support the steady-state assumption in our simulations, we present the time-averaged radial density profiles obtained over two time windows along with their associated standard deviations. As observed, the time-averaged density profiles remain the same (within their standard deviations). A similar qualitative evolution is observed with a higher number of GMs in the LC and HC regime, as well as in the cold-ion and Braginskii simulations.

 Regarding the computational cost, we remark that reaching a quasi-steady state requires a total of $\sim 41500$ CPU hours on the multi-core partition of the Piz Daint supercomputer, approximatively.

\begin{figure}
\centering
\includegraphics[scale =0.58]{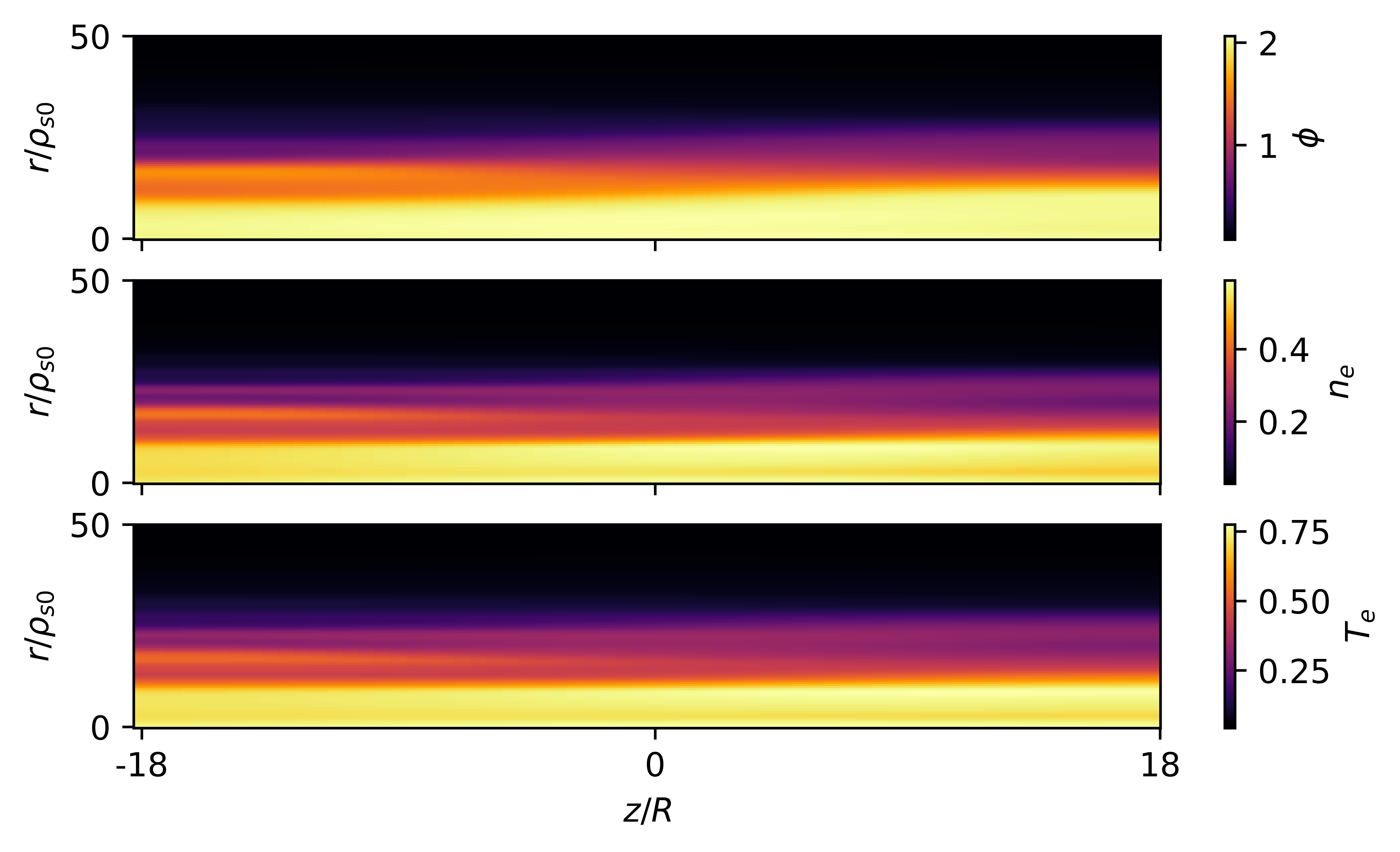}
\caption{Typical snapshots of the parallel turbulent structures of the electrostatic potential $\phi$ (top), electron density $n_e$ (middle), and electron temperature $T_e$ (bottom) using $(P,J) =(6,1)$ GMs in the HC regime during the quasi-steady state and as a function of $r$. We notice that $\phi \sim 3 T_e$. }
\label{fig:snapshotsxz}
\end{figure}

The dynamics in the direction parallel to the magnetic field is shown in \figref{fig:snapshotsxz} during the quasi-steady state. Instantaneous snapshots of the parallel turbulent structures of the electrostatic potential $\phi$, electron density $n_e$, and electron temperature $T_e$ reveal elongated structures. All quantities show weak parallel gradients $k_\parallel \simeq 0$, but have slightly larger values at the center ($z =0$) and decrease in amplitude slowly near the end plates (located at $z= - 18 R$ and at $z = 18 R$) due to the particle and energy losses caused by the sheath boundary conditions. Similar parallel structures are obtained when using a larger number of GMs. The turbulent structures observed in \figref{fig:snapshotsxz} are in qualitative agreement with previous fluid \cite{Rogers2010,Popovich2010,Fisher2015} and GK \cite{Shi2017,Pan2018} turbulent simulations of LAPD.

\begin{figure*}
\centering
\includegraphics[scale =0.7]{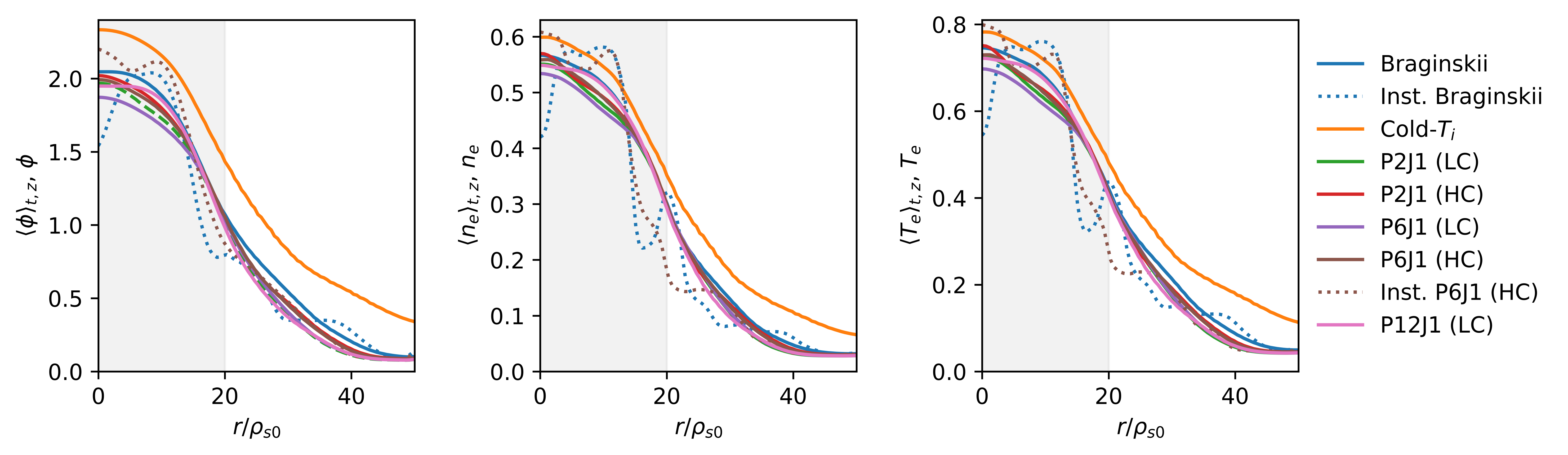}
\caption{Time-averaged profiles (solid colored lines) of the electrostatic potential $\left< \phi \right>_{t,z}$ (left), electron density $\left<  n_e \right>_{t,z}$ (center), and electron temperature $\left<T_e \right>_{t,z}$ (right) obtained in the GM simulations in the HC regime using $(P,J) = (2,1)$, $(6,1)$ and in the LC regime using $(P,J) = (2,1)$, $(6,1)$ and $(12,1)$, as a function of the radial coordinate $r$. The cold-ion and Braginskii simulations are also shown. The profiles are averaged over the region $ - 8  R \leq z \leq 8 R$. The instantaneous profiles are shown by the dotted colored lines for the Braginskii and $(P,J) = (6,1)$ GM in the HC regime simulations. The gray shaded areas represent the radial extent of the localized sources.}
\label{fig:profiles}
\end{figure*}

We now examine the time-averaged radial profiles. These profiles are obtained by averaging over a time window of $\sim 2$ ms during the quasi-steady state as well as over the central region of LAPD, i.e. $- 8 R \leq z \leq + 8R$ (or $- 4 \text{m} \lesssim z \lesssim 4 \text{m}$), a region commonly considered to present experimental data \cite{Carter2006} (a similar approach is used in previous GK simulations \cite{Shi2017,Pan2018}). More precisely, the averaged profile of the electrostatic potential $\phi$ is computed by defining $\left< \phi \right>_{t,z} = \int_{-\ell_z/2}^{+\ell_z/2} d z \int_0^{\tau}  dt  \phi /( \ell_z \tau ) $, with $\ell_z$ and $\tau$ are the $z$ and time length of the average windows, respectively. The results are shown in \figref{fig:profiles}, which displays the profiles of $\left< \phi \right>_{t,z}$, $\left<  n_e \right>_{t,z}$, and $\left<T_e \right>_{t,z}$ obtained in the GM simulations, using different numbers of GMs, in the cold-ion and the Braginskii simulations and as a function of the radius. Instantaneous profiles are also included for comparison.  

We note, first, that the plasma profiles extend beyond $r_s$ illustrating the broadening caused by the KH instability \cite{Rogers2010}. More precisely, the profiles are approximatively constant close to the center of the device ($r < r_s$) and far from the source region ($r > r_s$), showing a region of steep gradients near $r \sim r_s$, where the fluctuation level is large (see \secref{sec:turbulence}). Second, the time-averaged radial profiles from the GM simulations are very similar to the ones obtained from the Braginskii model. Third, no noticeable differences are found between the simulations in the LC and HC regimes and with different numbers of GMs. This suggests that ion kinetic effects may not significantly influence the predictions of the equilibrium (time-averaged) profiles in LAPD. On the other hand, the cold-ion model consistently predicts larger time-averaged radial profiles, while the gradients (not shown here) are of the same order as those obtained in the GM and Braginskii simulations. This difference between the cold-ion model and others cases with finite ion temperature is a consequence of the boundary conditions given in \eqref{eq:vparbc}, as discussed below. Fourth, the analysis of the instantaneous profiles (indicated by dotted lines in \figref{fig:profiles}) shows the existence of large perpendicular turbulent structures associated with the KH instability. Finally, we note that the time-averaged profiles obtained in \figref{fig:profiles} feature a similar behaviour of those obtained in previous fluid simulations \cite{Rogers2010} and GK simulations \cite{Shi2017,Pan2018}.

\begin{figure}
\centering
\includegraphics[scale =0.63]{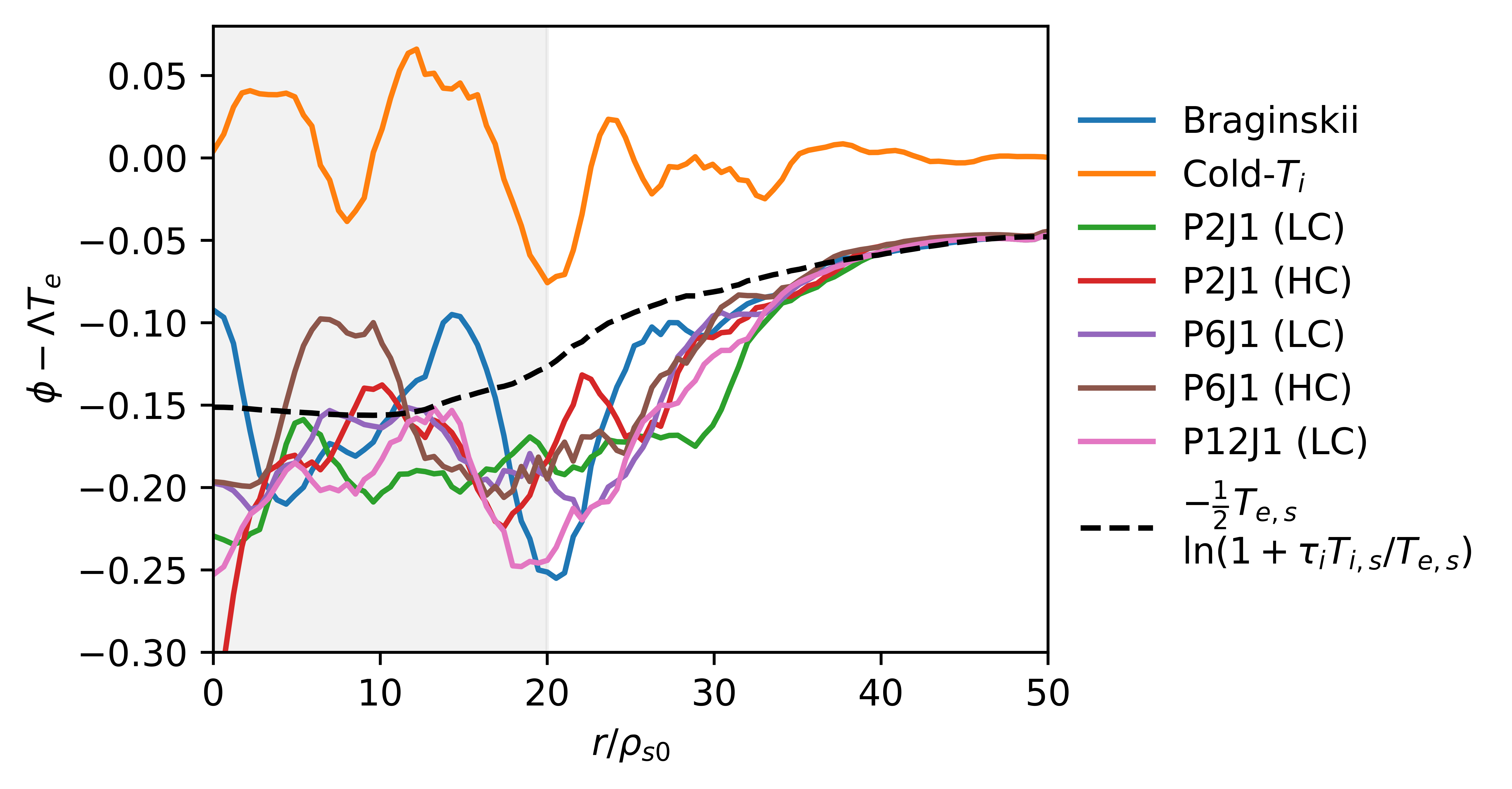}
\caption{Radial instantaneous profiles of the difference between $\phi$ and $\Lambda T_e$ at $z = 0 R$, in the LC and HC regimes for different number of GMs $(P,J)$ and at quasi-steady state. The results from the cold-ion and Braginskii simulations are also shown. The black dotted line shows the value of $- T_{e,s} \ln\left( 1 +  \tau_{i} T_{i, s} / T_{e,s}\right)/2$ in the Braginskii simulation at $z = 0R$ (similar values of obtained in the GM simulations). The gray shaded area ($r \le r_s$) represents the radial extent of the top-hat-like sources. For $r \gg r_s $, $\phi$ relaxes to $\Lambda T_e$. }
\label{fig:philambdaTe}
\end{figure}

We remark that the electrostatic potential profile $\phi$ follows approximatively the electron temperature $T_e$, as shown in \figref{fig:profiles}. Indeed, $\phi \sim \Lambda T_e$ is required to have comparable electron and ion outflows in steady-state, such that $U_{\|i} \sim U_{\| e}$ near the end plates, according to \eqref{eq:vparbc}. To verify that $ \phi \sim \Lambda T_e$ in our simulations, we evaluate the radial profile of the instantaneous difference, $\phi - \Lambda T_e$, taken at the center of the device ($z = 0 R$) for the GM (both the LC and HC regimes are considered), cold-ion and Braginskii simulations during the quasi-steady state. The results are shown in \figref{fig:philambdaTe}. We first observe that the GM and Braginskii simulations yield similar negative $\phi - \Lambda T_e$ values. On the other hand, $\phi - \Lambda T_e$ is roughly constant and approximatively vanishes for all radii in the cold-ion model. The deviations observed in $\phi - \Lambda T_e$ are the result of the boundary conditions specified in \eqref{eq:vparbc}. In fact, to achieve an equilibrium between parallel ion and electron flows, it is necessary that $\phi - \Lambda T_e \simeq -T_{e,s} \ln\left(1 + \tau_{i} T_{i, s} /T_{e,s}\right)/2$ (as indicated by the black dotted line in \figref{fig:philambdaTe}) when finite ion temperature is considered in the case of Braginskii and GM simulations. In contrast, this condition reduces to $\phi - \Lambda T_e \simeq 0$ in the cold-ion model. Consequently, due to the finite ion temperature dependence of the boundary condition \eqref{eq:vparibc}, the equilibrium electrostatic potential $\phi$ exhibits higher amplitude in the cold ion model, which is confirmed in the time-averaged profiles observed in \eqref{fig:profiles}. Finally, in all cases considered, $\phi - \Lambda T_e$ remains smaller than the values of $\phi$ and $\Lambda T_e $ ($ \phi - \Lambda T_e \sim 0.1$ for $r \lesssim r_s$ compared to $\phi \sim \Lambda T_e \sim 2$, see \figref{fig:snapshotsxz}).

\subsection{Turbulence analysis}
\label{sec:turbulence}

\begin{figure*}[htb]
\centering
\includegraphics[scale =0.62]{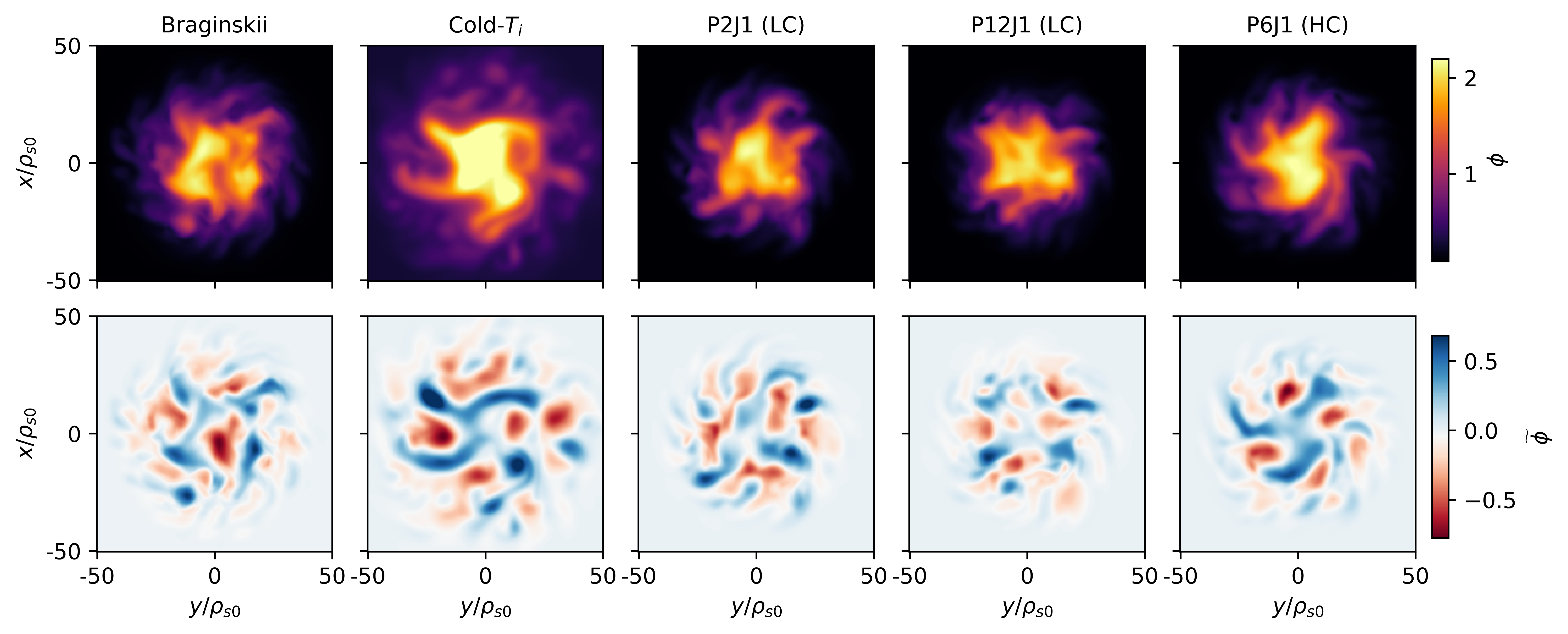}
\caption{Snapshots of the electrostatic potential $\phi$ (top) and of the fluctuations $\widetilde{ \phi} = \phi -\left< \phi \right>_t$ (bottom) taken at $z=0 R$ obtained in the Braginskii, cold-ion (cold-$T_i$), $(P,J) = (2,1)$, $ (12,1)$ in the LC regime, and $(P,J) = (6,1)$ in the HC regime simulations (from left to right).}
\label{fig:phisnapshots}
\end{figure*}

We now delve into the analysis of the turbulence properties, comparing the GM predictions with the Braginskii simulations. The instantaneous fluctuations are obtained by subtracting the time-averaged profiles from the full quantities, such that the fluctuation of, e.g., the electrostatic potential, $\widetilde{\phi}$, is defined by $\widetilde{\phi} =  \phi - \left< \phi \right>_t $, where $\left<\phi \right>_t =  \int_0^{\tau} dt  \phi / \tau $ denotes the time-averaged potential. Similar definitions for the other quantities are used. The top panels of \figref{fig:phisnapshots} show instantaneous snapshots of $\phi$ in the plane perpendicular at the center of the device $z=0 R$, while the bottom panels illustrate $\widetilde{\phi}$ snapshots. The Braginskii, cold-ion, and GM simulations with various $(P,J)$ are considered.

\begin{figure*}
\centering
\includegraphics[scale =0.62]{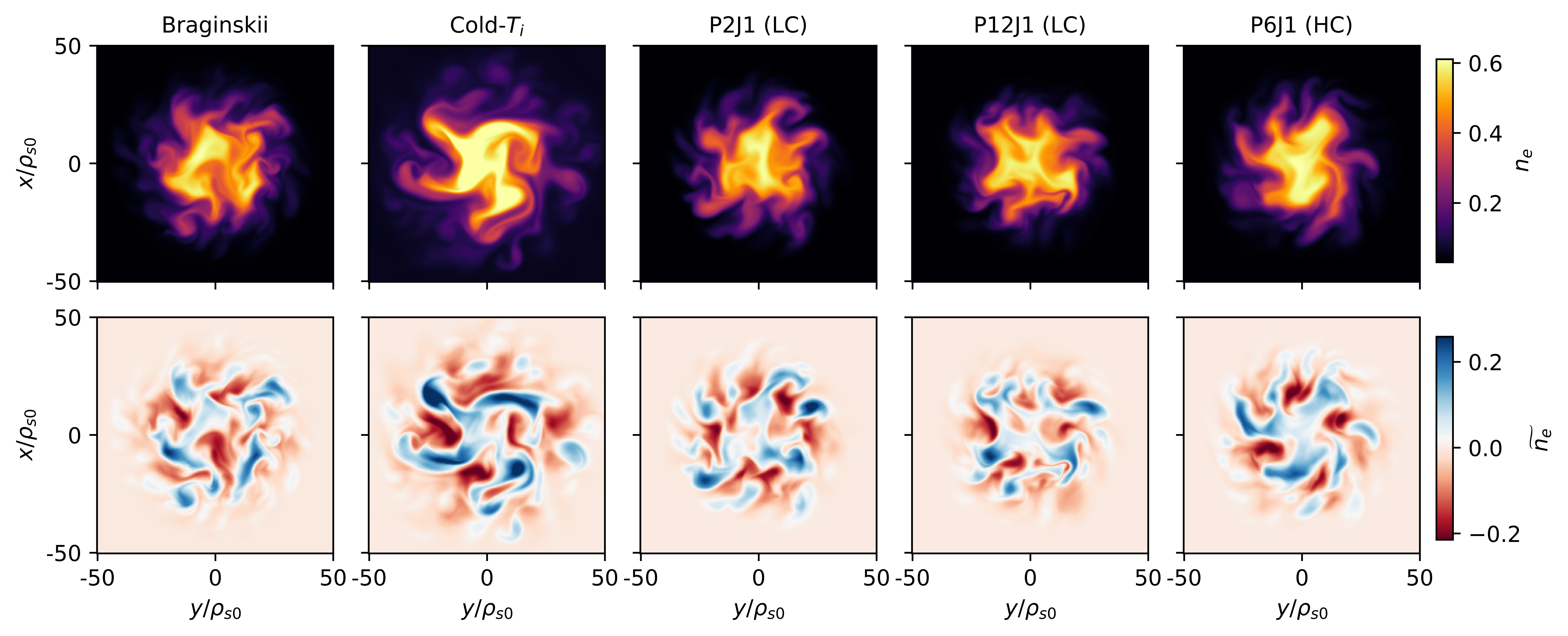}
\caption{Snapshots of the electron density $n_e$ (top) and of the fluctuation $\widetilde{n_e} = n_e - \left<n_e \right>_t$ (bottom), similar to \figref{fig:phisnapshots}.}
\label{fig:nisnapshots}
\end{figure*}

\begin{figure}
\centering
\includegraphics[scale =0.62]{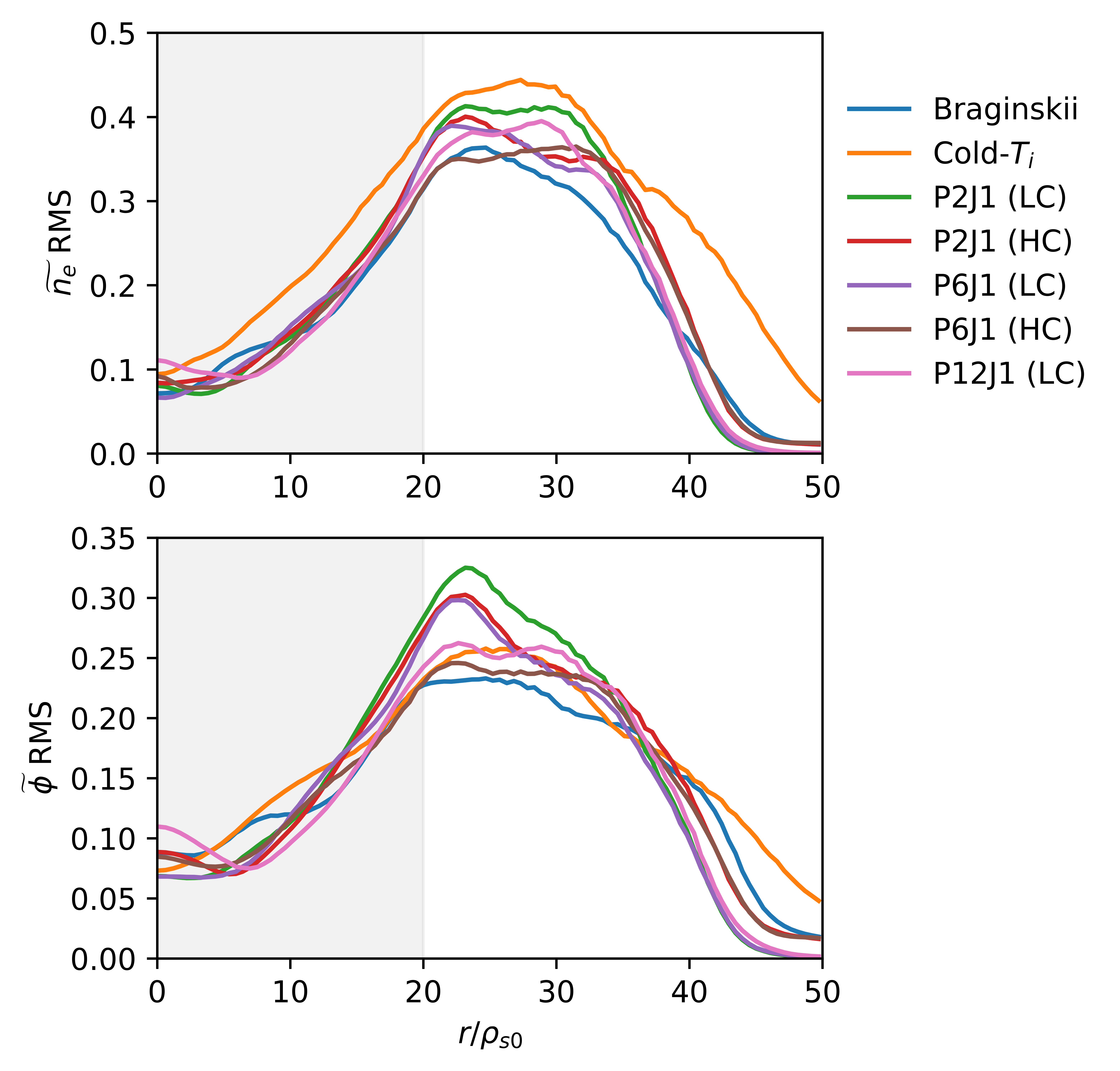}
\caption{RMS of the electron density $n_e$ (normalized to $\left< n_e \right>_t(r)$) (top) and electrostatic potential $\phi$ (normalized to $\left< \phi \right>_t(r)$) (bottom) fluctuations (computed from \figref{fig:nisnapshots} and \figref{fig:phisnapshots}) as a function of the radius $r$ in the case of the Braginskii, cold-ion, and GM simulations at $z=0R$. The shaded area indicates the radial extent of the sources.}
\label{fig:rms}
\end{figure}

We first observe that the fluctuations in the Braginskii model exhibits similar structures than those obtained in Ref. \onlinecite{Fisher2015}. In particular, the level of fluctuations is low at the center of the device and far from the source region, while it is large where the equilibrium gradient is steeper, in particular near $r \sim r_s$ (see \figref{fig:rms}). Notably, the $\widetilde{ \phi}$ snapshots reveal the presence of large amplitude structures propagating outwards. These observations hold for all the GM simulations, demonstrating qualitative agreement between the GM approach and the Braginskii model. While the fluctuations of the potential $\phi$ is not significantly affected by the number of GMs used in the simulation or by the collisionality regime, pointing out the fact that the KH instability (which drives turbulence) has a fluid nature, minor differences in the turbulent properties can still be observed. In fact, the use of a small number of GMs tends to produce slightly larger turbulent structures. This can be observed, for instance, by comparing the results of $(P,J) = (2,1)$ with the $(12,1)$ simulations in the LC regime. Finally, we observe that the cold-ion model produces the largest turbulent structures, which is consistent with the broad time-averaged profiles observed in \figref{fig:profiles}. The same observations apply to the snapshots of the electron density $n_e$ and its associated fluctuations $\widetilde{n_e}$, as shown in \figref{fig:nisnapshots}. Similar plots are obtained for $T_i $ and $T_e$, but not shown. Finally, it is worth noting that the fluctuations of the electrostatic potential $\widetilde{\phi}$ normalized to the electron temperature $T_e$ can be of the order of unity, as inferred from \eqref{fig:profiles} and \eqref{fig:phisnapshots}. In particular, we observe $\widetilde{\phi} / T_e \lesssim 1$ at all radii with, for example, $\widetilde{\phi} \sim 0.5$ and $T_e \sim 0.6$ for $r \lesssim r_s$.

\begin{figure}
\centering
\includegraphics[scale =0.63]{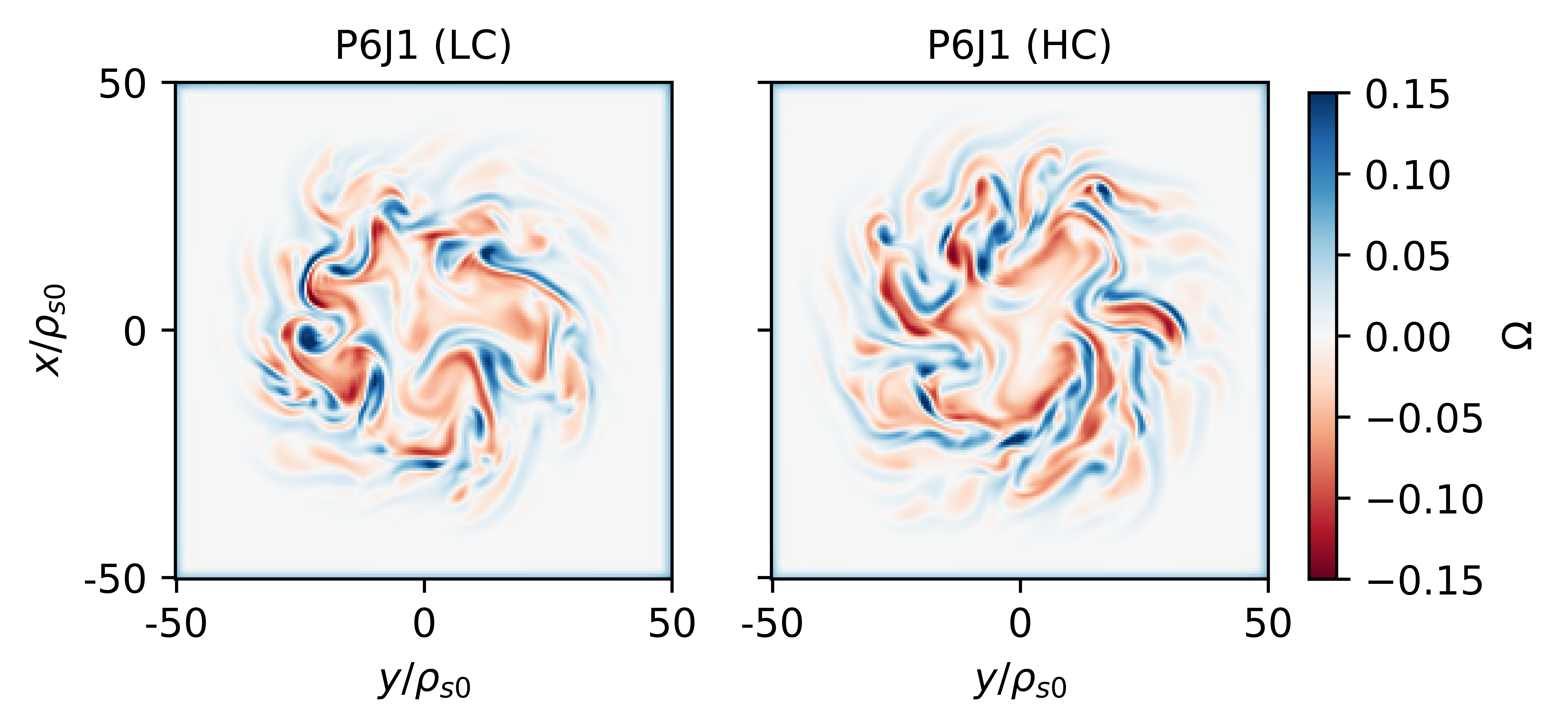}
\caption{Snapshots of the vorticity variable, $\Omega$, obtained in the GM simulations with $(P,J) = (6,1)$ in the LC (left) and HC (right) regimes at $z =0R$.}
\label{fig:omega}
\end{figure}

In \figref{fig:omega}, we present snapshots of the vorticity for the $\bm{E} \times \bm{B}$ flow, which is described by the vorticity variable $\Omega = \grad_\perp^2 \phi$, defined in \eqref{eq:vorticityboussinesq2}, at the center of the device ($z=0R$) at steady-state obtained in the case of the GM simulations with $(P,J) = (6,1)$ in the LC and HC regimes. It is observed that the vorticity exhibits large perpendicular structures, i.e. $k_\perp \rho_{s0} \ll 1$, driven by the KH instability, which develop in the region of steep equilibrium gradients near $r \sim r_s$. These structures have slightly smaller amplitude away from $r_s$ in all cases. Additionally, no clear differences are observed between the LC and HC regimes. Similar plots are obtained with the Braginskii and GM simulations. Finally, we note that further numerical investigations are necessary to verify the numerical resolution of the smallest perpendicular scale revealed by \figref{fig:omega}.

We now proceed to analyze the root mean square (RMS) of the fluctuations, defined as $\sqrt{\left< \widetilde{n_e}^2 \right>_t}$ in the case of the electron density fluctuations $\widetilde{ n_e}$ (and similarly for the other quantities). \figref{fig:rms} displays the RMS of the electron density, $n_e$, and the electrostatic potential, $\phi$, fluctuations plotted as a function of the radius. The data are computed at $z = 0 R$ and normalized to $\left<n_e \right>_t(r)$ (and to $\left<\phi \right>_t(r)$) \cite{Friedman2012,Shi2017}. We find that the RMS values of the density displayed in \figref{fig:rms} exhibit a qualitatively similar behaviour to those obtained in previous fluid simulations \cite{Fisher2015, Ross2019} and GK simulations \cite{Shi2017}. Consistent with the observations made in \figref{fig:nisnapshots}, the RMS values reach their maximum when the gradients are most pronounced near $r \sim r_s$. For $r \lesssim r_s$ and for $r \gtrsim r_s$ (where the gradients are smaller), the RMS values decrease because of the absence of the instability drive. Using a low number of GMs or considering the LC regime results in slightly larger RMS values (in particular of $\widetilde{\phi}$). Overall, this indicates that the level of fluctuations in the steep gradient region can be sensitive to the number of GMs used in the simulations. Finally, we note that the GM simulation with $(P,J) = (6,1)$ in the HC regime is the closest to the Braginskii predictions, and the largest RMS values (especially for $\widetilde{n_e}$) are obtained in the cold-ion model.

\begin{figure}
\centering
\includegraphics[scale =0.62]{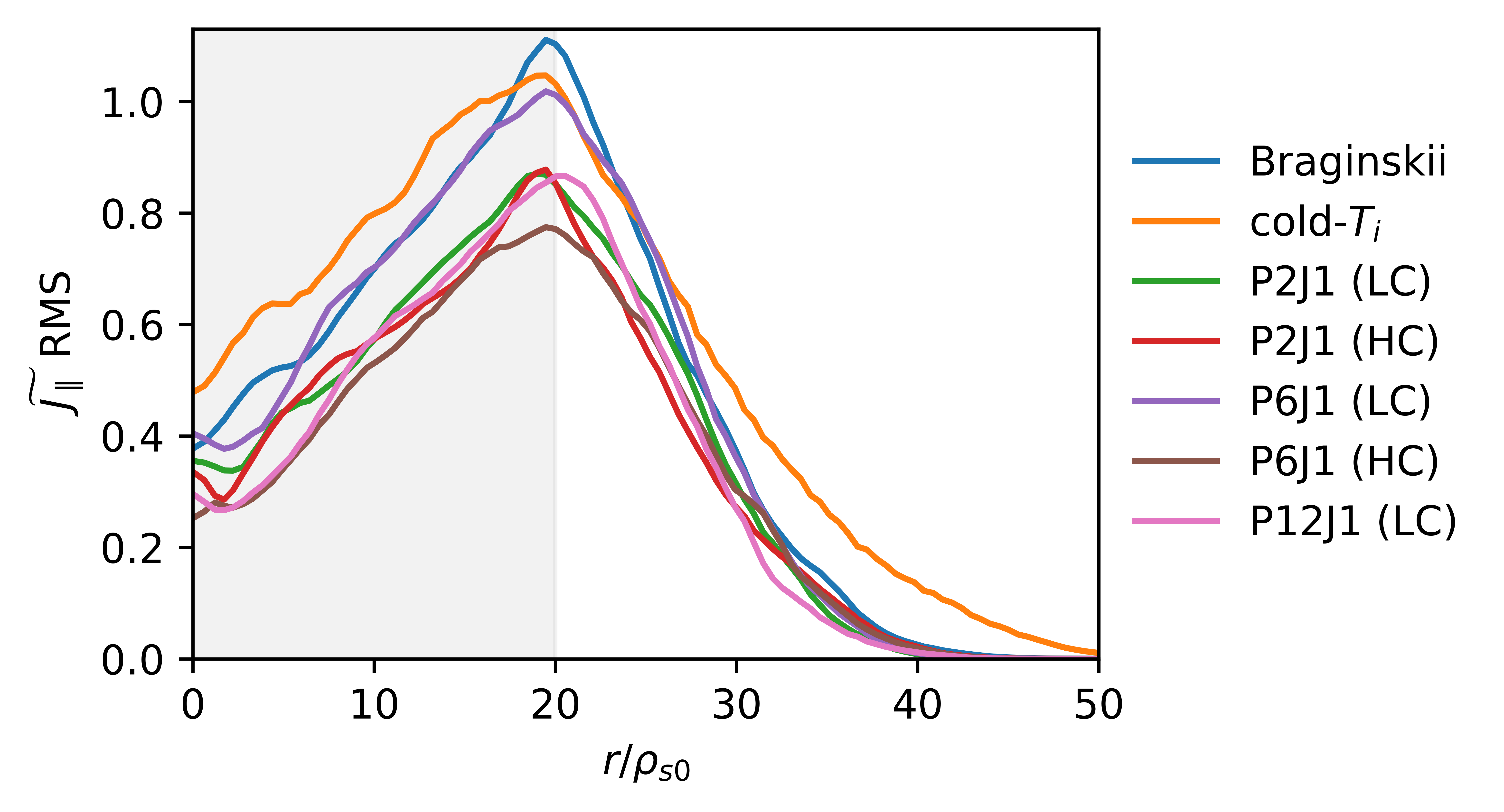}
\caption{RMS of the total parallel current $J_{\parallel}$ (normalized to the maximum of $\left< J_{\parallel} \right>_t(r)$) as a function of the radius at the sheath entrance near $z = - 18 R$.}
\label{fig:rmsjpar}
\end{figure}

We compare the RMS of the parallel electrical current $J_\parallel$ measured at the sheath entrance located at $z = - 18 R$. The results are shown in \figref{fig:rmsjpar} as a function of the radius and normalized to the maximum of $\left< J_\parallel \right>_t(r)$. It is clearly observed that the boundary conditions imposed on the electron and ion parallel velocities allow for the parallel current to fluctuate. This is in contrast to the case of logical sheath boundary condition, where $J_\parallel = 0$ is imposed everywhere \cite{Parker1992}. We remark that larger fluctuations of $J_\parallel$ are obtained in the Braginskii simulations, while the largest RMS is observed in the case of the cold-ion model, a consequence of the finite ion temperature effect in the boundary conditions retained in the former (see \eqref{eq:vparbc}) and ignored in the later.

\begin{figure}
\centering
\includegraphics[scale =0.64]{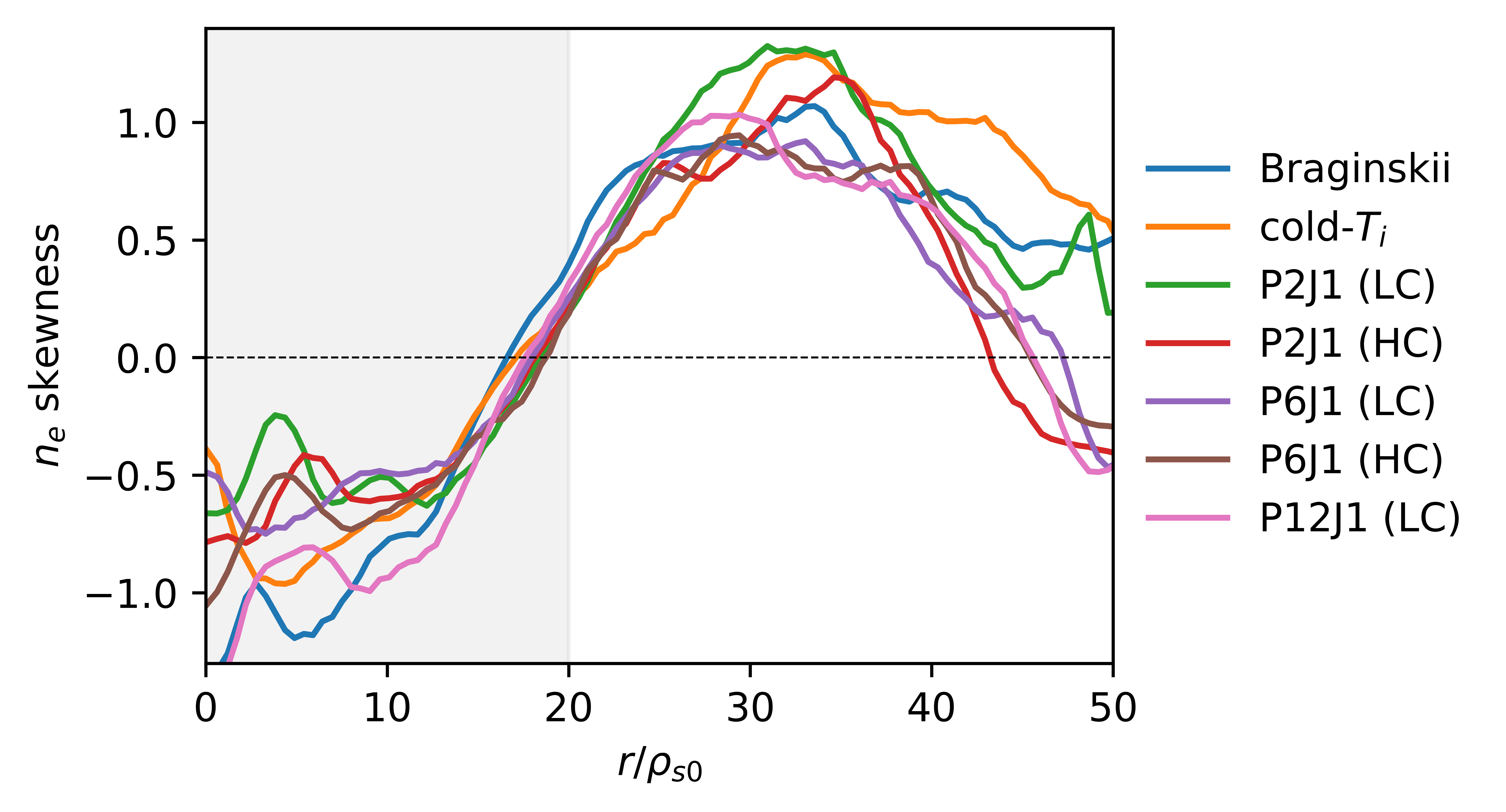}
\caption{Skewness of the electron density $n_e$ as a function of $r$ at $z=0 R$. The gray shaded area represents the radial extent of the sources.}
\label{fig:skewness}
\end{figure}

We now turn our attention to the skewness of the ion density fluctuations, which is defined as the third normalized moment of the electron density fluctuation, that is $\left< \widetilde{n_e}^3 \right>_t  / \left< \widetilde{n_e}^2 \right>_t^{3/2}$. The skewness of the density is often used to characterize the dominance of plasma holes and blobs, associated with negative and positive skewness respectively \cite{Friedman2012,Fisher2015,Ross2019}. \figref{fig:skewness} shows the skewness of the electron density $n_e$. In all cases, the skewness is negative for $r \lesssim r_s$, indicating the presence of density holes in the region where the plasma source is present. On the other hand, in the region where $r \gtrsim r_s$, the skewness is positive. The sign and amplitude of the skewness shown in \figref{fig:skewness} have similar qualitative behaviour as in previous fluid \cite{Ross2019} and GK \cite{Shi2017,Pan2018} simulations. In particular, the values obtained in the GM simulations are of the same order to those observed in the Braginskii case, albeit slightly smaller. Overall, the present turbulent analysis demonstrates that the full-F GM approach is in qualitative agreement with the Braginskii model, which is also employed in previous numerical investigations \cite{Fisher2015,Ross2019} and validated with experimental data \cite{Friedman2012}. We note that a more comprehensive analysis of turbulence, including aspects such as particle and energy transport, is necessary to evaluate the impact of velocity-space resolution and collisions (along with collision operator models) in greater detail. This analysis is deferred to future investigations, which would involve extending the present work to include FLR effects and removing the Boussinesq approximation, important for accurate turbulence transport predictions.

\subsection{Ion distribution function at quasi-steady state}
\label{sec:vsp}

\begin{figure*}
\centering
\includegraphics[scale =0.8]{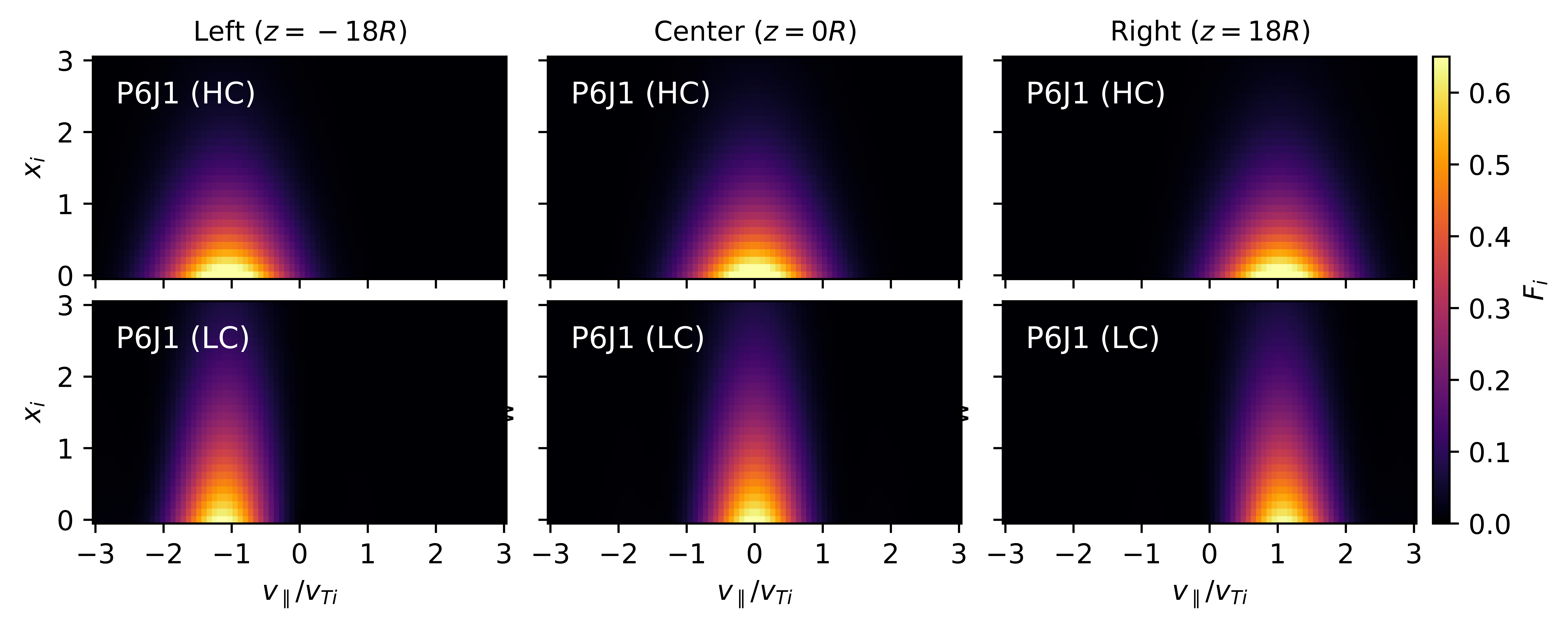}
\caption{Quasi-steady state ion distribution function, $F_i$, plotted as a function of $(v_\parallel / v_{Ti},x_i)$, obtained using $(P,J)= (6,1)$ GMs in the HC regime (top) and LC regime (bottom). $F_i$ is computed at the $z=-18 R$ (left), $z=0 R$ (center), and $z =18 R$ (right) at the center in the perpendicular plane ($x=y=0$).}
\label{fig:vsp}
\end{figure*}

\begin{figure}
\centering
\includegraphics[scale =0.7]{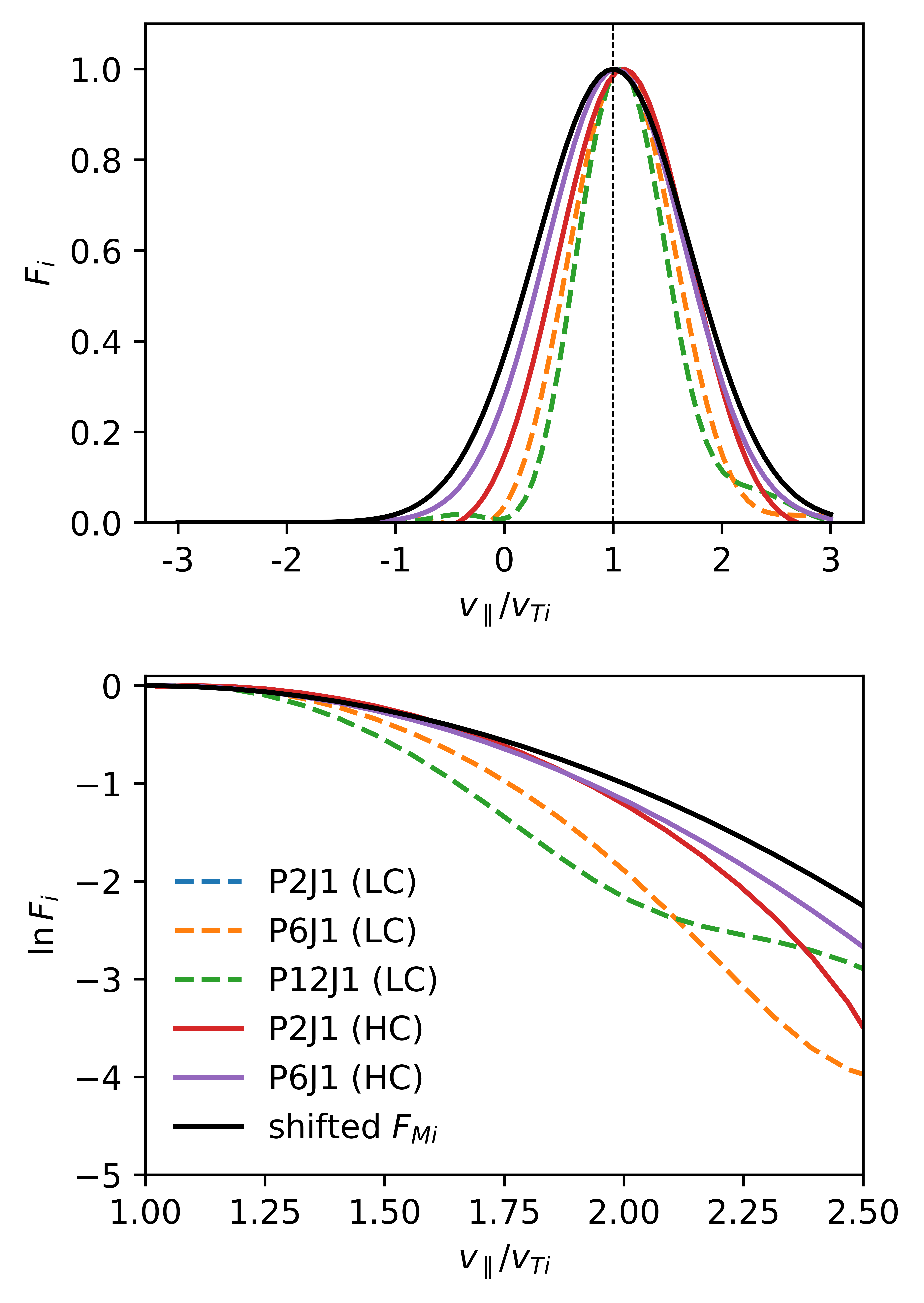}
\caption{Cuts of the ion distribution function $F_i$ (normalized to its maximum) at $x_i = 0$ at the right sheath entrance near $z = 18 R$ and $x=y=0$ (top) for different $(P,J)$ in the HC (solid lines) and LC (dotted lines) regimes, and close view over the region $v_\parallel  /v_{Ti} \in \left[ 1,2.5 \right]$ (bottom). The solid black line represents a Maxwellian distribution function shifted by the ion sound speed $c_s$ (see \eqref{eq:fmi}). A similar plot is obtained at the left sheath entrance ($z = -18 R$).}
\label{fig:vspslices}
\end{figure}

We now investigate the features of the ion distribution function $F_i$ in velocity-space. To obtain the full-F ion distribution function, $F_i$, from the GM simulations, we use the expansion in \eqref{eq:fullFi}, truncated to a finite number of GMs, and we compute it as a function of $x_i$ and the unshifted parallel coordinate, given by $v_\parallel / v_{Ti} = s_{\| i} +  \sqrt{2 \tau_i} U_{\| i }$. Also for this analysis, we consider the quasi-steady period. \Figref{fig:vsp} shows $F_i$ obtained from the $(P,J) = (6,1)$ simulations in the HC and LC regime at the center of the machine ($z=0R$) and at the sheath entrances, $z=- 18R$ and $z=18R$. At the two sheath entrances, $F_i$ is centered around the ion parallel velocity, $U_{\ i} =\pm c_s$ respectively, a consequence of the Bohm sheath boundary conditions given in \eqref{eq:vparbc}. On the other hand, $F_i$ is centered around $v_\parallel \simeq 0$ at $z = 0R$, where $U_{\| i} \simeq 0 $. The absence of fine velocity-space structures in \figref{fig:vsp} in both the HC and LC regimes confirms the weak dependence of turbulence properties on the number of GMs, as reported in \secref{sec:turbulence}. In fact, fine velocity-space structures in $F_i$ are not expected to play a significant role in LAPD due to the absence of magnetic drifts, trapped particles, of the high-collisional regime, and the fluid nature of the KH-dominated turbulent transport. Nevertheless, we remark that strong velocity-space gradients can appear near the sheath entrance in the electron distribution function (see, e.g., Ref. \onlinecite{Pan2018}). The Maxwellian shape of $F_i$ observed in our simulations is also consistent with the fact that both two-fluid Braginskii \cite{Rogers2010,Ross2019} and GK \cite{Shi2017,Pan2018} simulations of LAPD yield qualitatively similar results.

\Figref{fig:vspslices} shows the ion distribution function at the sheath entrance ($z = 18 R$ and $x = y = 0$) for $x_i =0$, in the LC and HC regimes and for different values of $(P,J)$. We first observe that the bulk region of $F_i$ (near $v_\parallel / v_{Ti} \sim 1$) is well approximated by a shifted Maxwellian. However, deviations from the Maxwellian distribution function are noticeable in the tails of $F_i$ in the LC regime. These deviations become pronounced as $(P,J)$ increases (e.g., from $(6,1)$ to $(12,1)$) which indicates that $F_i$ is not sufficiently resolved in the LC regime. Finally, we remark that collisional effects tend to widen $F_i$ due to the collisional parallel velocity-space diffusion present in the nonlinear Dougherty operator. We note that further convergence studies are required to fully evaluate the velocity-space resolution in \figref{fig:vspslices} by considering a larger number of GMs than those considered in the present work, in particular in the LC regime.

The GM hierarchy equation in \eqref{eq:ionhierarchy} does not enforce the positivity of the ion distribution function when evolving a finite number of GMs. For instance, the use of $v_\parallel / v_{Ti}$ as an argument in the Hermite polynomials, $H_p$, in \eqref{eq:fullFi} would compromise the convergence properties of the GM approach, with respect to the use of $v_\parallel / v_{Ti}$, leading to simulations that show unphysical distribution functions with negative values when the same number of GMs are considered, as in the simulations presented here (\figref{fig:vsp} would show negative values). In fact, if the unshifted GMs $\N_{v_\parallel}^{pj}$, defined with respect to $v_\parallel / v_{Ti}$ as the argument of $H_p$, i.e.

\begin{align} \label{eq:Nipjdef2}
\mathcal{N}^{pj}_{v_\parallel}  =    2 \pi \int_{- \infty}^\infty d v_\parallel \int_{0}^\infty d \mu \frac{B}{m_i } F_i \frac{H_p \left( \frac{v_\parallel }{ v_{Ti}}\right) L_j(x_i)}{\sqrt{2^p p!}},
\end{align}
\\
 are used to expand $F_i$, it is found that $\N_{v_\parallel}^{pj} \neq 0$ for $(p,j) > 0$, even when $F_i$ is a Maxwellian distribution function centered at $U_{\| i} \neq 0 $. Indeed, using \eqref{eq:Nipjdef2}, one derives the analytical expression of the unshifted GMs for $F_i = F_{Mi}$, 

\begin{align}
\N_{v_\parallel}^{pj} & =   \frac{\delta_j^0}{\sqrt{\pi}} \int_{- \infty}^{\infty} d \left( \frac{v_\parallel}{v_{Ti}} \right) e^{- (v_\parallel / v_{Ti} - U_{\| i} / (\sqrt{2 \tau_i } ))^2 } \frac{H_p\left(  \frac{v_\parallel}{v_{Ti}}\right) }{\sqrt{2^p p!}}   \nonumber  \\
& =  \sqrt{\frac{2^p}{p!}} \left(\frac{U_{\| i}}{\sqrt{2 \tau_i } } \right)^p \delta_j^0,\label{eq:npjunshifted}
\end{align}
\\
where $U_{\| i}$ is normalized to $c_{s0}$. While the amplitude of the unshifted GM decreases rapidly in the presence of subsonic ion flow, $U_{\| i} \ll 1$, the decrease of the amplitude with $p$ is slower in the presence of sonic flows, such that $\N^{pj}_{v_\parallel} \sim \sqrt{2^p / p!}$.  

\subsection{GM spectrum at quasi-steady sate}
\label{subsec:GMspectrum}

\begin{figure}
\centering
\includegraphics[scale =0.7]{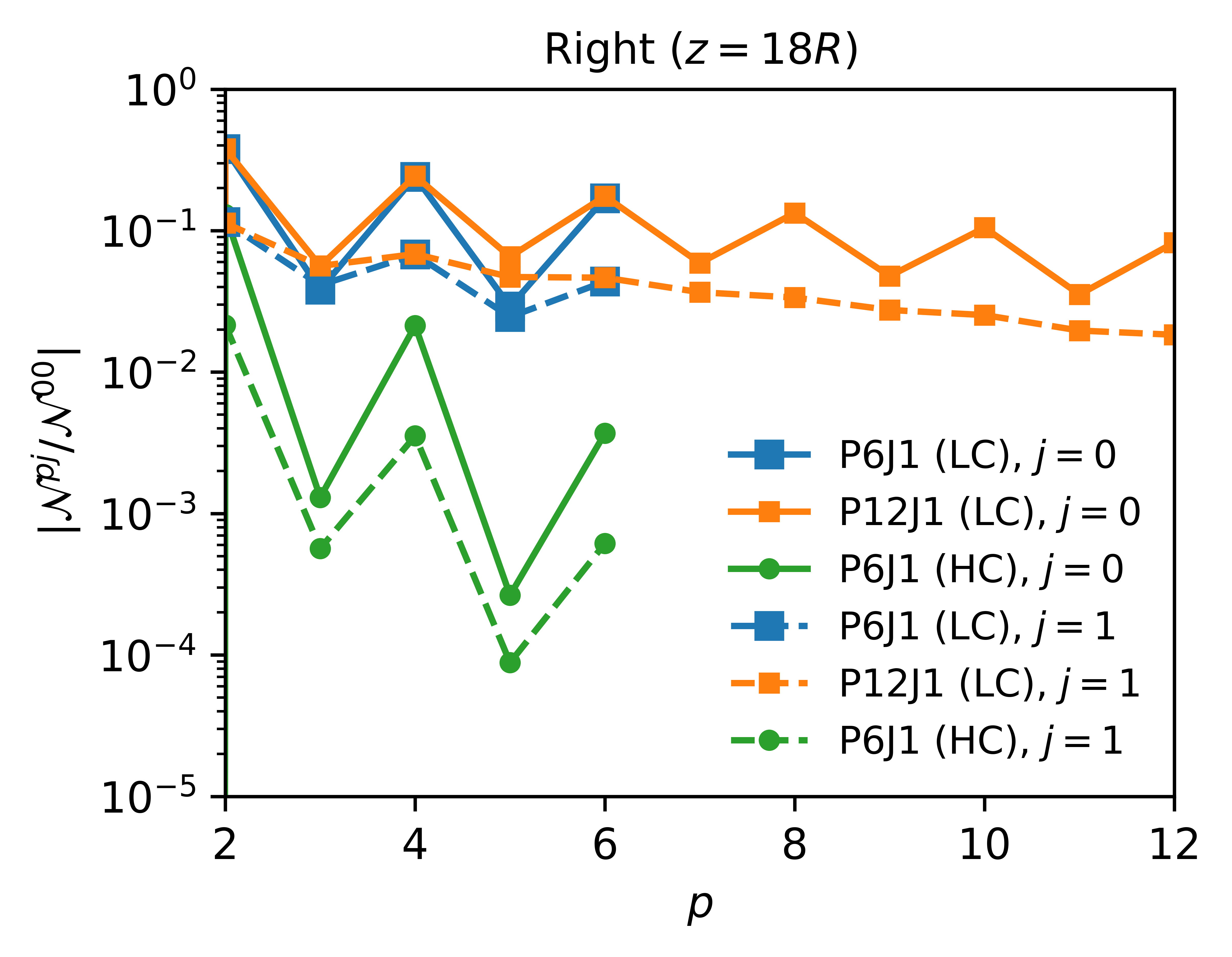}
\caption{Absolute value of the GMs $\N^{pj}$ (normalized to $\N^{00}$) associated with the distribution functions shown in \figref{fig:vspslices}, plotted on a logarithmic scale as a function of $p \geq 2$ (solid lines for $j=0$ and dashed lines for $j=1$) obtained from different $(P,J)$ simulations in the LC (square colored markers) and HC (circle colored markers) regime at $z = 18 R$ and $x=y=0$. The amplitude of the GM decreases with $p$ in all cases and it is faster in the HC regime.}
\label{fig:gmspectrum}
\end{figure}

To better assess the velocity-space representation of $F_i$ in our simulations, we plot the absolute value of the GMs, $\N^{pj}$ (normalized to $\N^{00}$), at the sheath entrance of the device, $z=18 R$ and $r =0$, in \figref{fig:gmspectrum}. We consider the GMs associated with the distribution functions displayed in \figref{fig:vspslices}. As it can be clearly observed, the amplitude of the GMs decays faster in the HC regime than in the LC regime. In fact, extrapolating from linear theory in slab geometry \cite{Frei2022}, the amplitude of the GMs is expected to follow $ \left| \N^{pj} \right| \sim \exp\left[ - c_p(\nu_i) p^{\alpha} \right]$ in the $p \gg 1$ limit (with $c_p(\nu_i)$ a positive coefficient that increases with collisionality, $\nu_i$ and $\alpha> 0$). While this estimate has been verified in linear studies and remains to be proven valid in full-F turbulent simulations, it provides us with a useful proxy for the expected decay of the GMs in the presence of collisions.

The results show that $P > 12$ would be required to ensure that $F_i$ is sufficiently well resolved in the LC regime. Since the smaller relevance of the LC regime for the experimental parameters of LAPD, simulations with $P > 12$ have not been carried out in the present work to properly assess the velocity-space resolution of $F_i$. On the other hand, the contributions from $\N^{p0}$ with $p \gtrsim 4$ are negligible in the HC regime, thereby justifying the closure by truncation for $P \gtrsim 4$. We also notice that $\N^{10} =0$ in all cases, as a consequence of \eqref{eq:N10}. Finally, we note that the amplitude of the low-order GMs is not sensitive to $P$, as shown in \figref{fig:gmspectrum}. More precisely, the low-order GMs for $(P,J) = (6,1)$ strongly resemble the ones of the $(P,J) = (12,1)$ simulation in the LC case. This holds true also in the HC regime, for instance, by comparing the $(P,J) = (6,1)$ and $(P,J) = (2,1)$ simulations.

We also show in \figref{fig:gmspectrum} that the amplitudes of the $j=1$ GMs are smaller than the ones with $j=0$. Numerical experiments (not shown) have also revealed that GMs with $j>1$ have a negligible amplitude due to the absence of source for $j >1$ and of FLR effects. This suggests (in addition to the similar results obtained in \secref{sec:turbulence} with different $(P,J)$) that full-F turbulent calculations using the GM approach are less sensitive to the values of $P$ and $J$ than linear computations \cite{Jorge2019epw,Frei2022}, where applying a closure by truncation at low $P$ and $J$ can introduce spurious artifacts \cite{Frei2023}. Otherwise, \figref{fig:gmspectrum} reveals that the slightly larger RMS values depicted in \figref{fig:rms} (e.g., $(P,J) = (6,1)$ in the LC and $(P,J) = (2,1)$ in the HC regime) correspond to cases where the GM representation of $F_i$ is unresolved. Additional investigations are required to verify the effect of closure in the presence of kinetic effects such as trapped particles and magnetic drifts, which are absent in LAPD, and of FLR effects, which are neglected in our model.

\begin{figure*}
\centering
\includegraphics[scale =0.72]{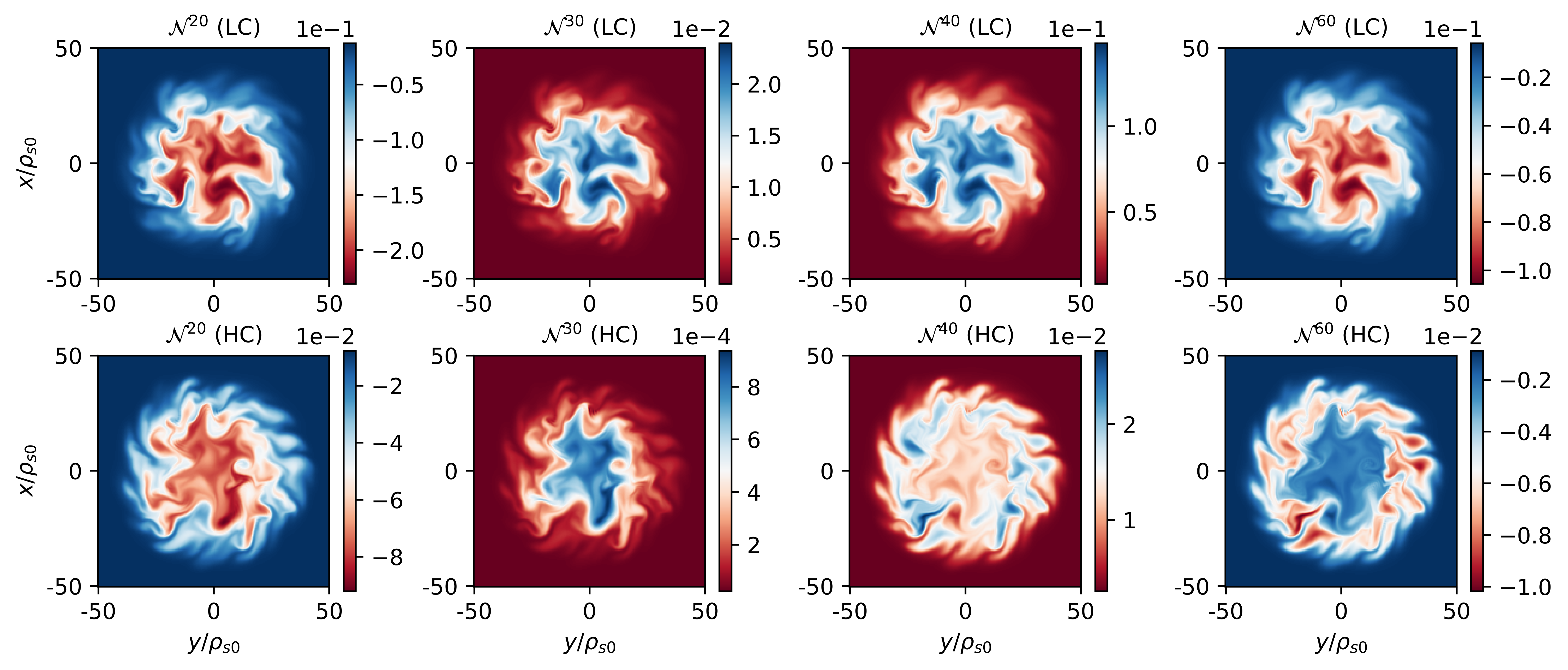}
\caption{Instantaneous snapshots of the GMs $\N^{pj}$, with $p = 2,3,4,6$ and $j=0$, in the $(x,y)$ perpendicular plane at the sheath entrance located at $z = 18 R$ from the $(P,J) = (6,1)$ simulations in the LC (top) and HC (bottom) regime. The large perpendicular structures are caused by the KH instability. }
\label{fig:npj}
\end{figure*}

Finally, \figref{fig:npj} presents snapshots of the GMs for different values of $p$ in the perpendicular plane obtained for the $(P,J) = (6,1)$ simulations in the LC and HC regimes. It is also visible that the turbulent structures are dominated by a long-wavelength perpendicular fluctuations for all values of $p$, which are driven by the KH instability. Indeed, the amplitude of these turbulent structures is strongly reduced if the nonlinear drive of the KH instability is artificially removed from our simulations. The decay of the amplitude of the turbulent structures due to collisions and with increasing $p$ is also evident. 
 
\section{Conclusions}
\label{sec:conclusion}

In this work, we present the first full-F turbulent simulations based on the GM approach using an arbitrary number of moments in a linear plasma device configuration with open straight field lines, such as LAPD. Neglecting FLR effects, we consider an electrostatic and long-wavelength ion GK model for the full ion distribution function $F_i$, coupled to the electron Braginskii fluid model for the electron density $n_e$, parallel velocity $U_{\| e}$, and temperature $T_e$. The ion GK model is solved by deriving a full-F ion GM hierarchy equation, based on the Hermite-Laguerre polynomial expansion of $F_i$. In particular, a velocity-space coordinate centered at the local fluid ion parallel velocity is used to expand $F_i$, which ensures good convergence properties of the Hermite expansion in the presence of sonic ion flows. The GM hierarchy equation we consider is equivalent to the electrostatic and Boussinesq limits of the GK moment model for the boundary region derived in Ref. \onlinecite{Frei2020} with vanishing FLR effects. To account for the parallel losses at the end plates, Bohm sheath boundary conditions are used, equivalent to those previously used in LAPD fluid simulations \cite{Rogers2010}. A nonlinear ion-ion Dougherty collision operator is also considered. The ion GM hierarchy equation with the fluid electron model is implemented in a numerical code, allowing us to perform the first full-F turbulent calculations based on a moment approach with a flexible number of GMs. This is in contrast to previous full-F gyrofluid simulations where the number of moments is fixed \cite{Wiesenberger2019,Galassi2022,Wiesenberger2022}.

We present the simulations of a linear device using LAPD physical parameters based on a Helium plasma \cite{Rogers2010} and a first-of-the-kind comparison with the two-fluid Braginskii model. Several nonlinear simulations are performed using a different number of Hermite and Laguerre GMs in a low and high-collisional ion regime. Overall, a qualitative agreement on the time-averaged radial profiles between the Braginskii model and the GM approach is observed. This is expected from our analysis which shows that turbulence is dominated by the long perpendicular wavelength and $k_\parallel \simeq 0$ Kelvin-Helmoltz instability of fluid nature.

 The RMS and skewness of the fluctuations in the GM simulations also agree with the ones previously obtained in fluid \cite{Rogers2010,Fisher2015} and GK \cite{Shi2017,Pan2018} simulations of LAPD. In particular, we find that the RMS values are often larger than the ones predicted by the Braginskii model, if the number of GMs is not sufficient to properly resolve the ion distribution function. The largest RMS values are observed with the cold-ion reduced model (with a difference up to $\sim 20 \%$ with respect to the Braginskii model), while the results closest to the one of the Braginskii model are obtained if collisions are introduced in the GM approach with a sufficient number of GMs, in this case $(P,J) = (6,1)$. Overall, collisions reduce slightly the turbulent fluctuations level, but they do not significantly alter the observed turbulent regimes. At the same time, the analysis of the ion distribution function $F_i$ reveals that collisions damp the amplitudes of the GMs, thereby allowing for a reduction in the number of GMs required in the simulations (typically from $(P,J) \sim (12,1)$ in the low collisional regime to $(P,J) \sim (6,1)$ in the high-collisional regime of LAPD). 

We note that although the present work allows the simulation of LAPD with a flexible number of GMs, unlike previous full-F gyrofluid simulations\cite{Wiesenberger2019,Galassi2022,Wiesenberger2022}, improvements in our physical model are required to increase the fidelity of our simulations. For example, the inclusion of FLR effects \cite{Held2020,Held2022} is necessary to study small scale fluctuations. A more detailed study of the role of FLR effects is the subject of future work. Incorporating FLR effects with a flexible number of GMs necessitates conducting comprehensive numerical investigations to evaluate the impact of the numerical and velocity-space resolution on turbulent structures, in particular at small scales, supported by the examination of conservation laws, an aspect not explored in the present work. In addition, a more accurate description of the role of collisions requires the implementation of a nonlinear collision operator model with increasing physical fidelity, such as the nonlinear Coulomb operator\cite{Jorge2019}. Proper sheath boundary conditions for the GM hierarchy equation, which extend the simplified Bohm sheath boundary condition used here (see \eqref{eq:vparbc}), can improve the reliability of our simulations. These boundary conditions can be obtained by a procedure similar to that described in Ref. \onlinecite{Loizu2012}. Finally, we note that the implementation of a kinetic electron description is also essential to perform high-fidelity LAPD simulations, since fast and less collisional electrons (with $T_e \sim 15 $ eV) are emitted by pulsed plasma discharges in experiments \cite{Gekelman2016}. Furthermore, kinetic electrons are important in setting the sheath boundary conditions where electrons are reflected due to the potential drop, giving rise to strong speed-space gradients in the electron distribution function \cite{Pan2018}. 

More generally, the present work represents a step towards the development of future full-F turbulent simulations of the boundary region of fusion devices using the GM approach with a flexible number of GMs, which may provide an ideal flexible tool to capture kinetic and collisional effects at the desired level of accuracy.

\section*{Acknowledgement}

The authors acknowledge helpful discussions with Alessandro Geraldini and Stephan Brunner. This work has been carried out within the framework of the EUROfusion Consortium, via the Euratom Research and Training Programme (Grant Agreement No 101052200 — EUROfusion) and funded by the Swiss State Secretariat for Education, Research and Innovation (SERI). Views and opinions expressed are however those of the author(s) only and do not necessarily reflect those of the European Union, the European Commission, or SERI. Neither the European Union nor the European Commission nor SERI can be held responsible for them. The simulations presented herein were carried out in part on the CINECA Marconi supercomputer under the TSVVT422 project and in part at CSCS (Swiss National Supercomputing Center). This work was supported in part by the Swiss National Science Foundation.

\appendix

\section{High-collisional limit of ion temperature equation}
\label{appendixA}

In this appendix, we derive the Braginskii equation for the ion temperature $T_i$, given \eqref{eq:tibraginskii}, from the ion GM hierarchy equation provided in \eqref{eq:ionhierarchy}. This allows us to elucidate the relationship between the GM approach and the Braginskii model, particularly in the high-collisional regime, and to justify the terms on the right-hand side of \eqref{eq:tibraginskii}.

First, we note that since our model neglects FLR effects, the gyrocenter quantities are assumed to be equal to the particle fluid quantities. Hence, the time evolution of the ion temperature $T_i$ can be expressed in terms of GMs using the definitions of the gyrocenter fluid quantities in \eqref{eq:fluid2GMs}. Evaluating the time derivative of $T_i$ yields 

\begin{align} \label{eq:dtempidt}
N_i \partial_t T_i  = \frac{\sqrt{2}}{3} \partial_t \N^{20}   -  \frac{2 }{3}   \partial_t \N^{01}   + ( 1 -T_i )  \partial_t N_i.
\end{align}
\\
The time derivatives of the GMs contained in \eqref{eq:dtempidt} can be evaluated using the ion GM hierarchy equation given in \eqref{eq:ionhierarchy} with $(p,j) = (2,0)$ and $(0,1)$. We obtain 

\begin{subequations} \label{eq:dn20dtanddn01dt}
\begin{align}
\partial_t \N^{20} & = - \sqrt{\tau_i}  \partial_z
\sqrt{3}\N^{3 0}  -  \partial_z  \left( U_{\| i}  \N^{ 20} \right)  - \frac{1}{\rho_*} \poissonbracket{\phi}{\N^{ 20}} \nonumber \\ 
&   -  \left(  2\N^{20} + \sqrt{2} \N^{00}  \right)  \partial_z U_{\| i}  +  \C^{20}_i  +  S_\mathcal{E}^{20},
 \end{align}
\begin{align}
\partial_t \N^{01} & =  - \sqrt{\tau_i}  \partial_z
\N^{11} -  \partial_z  \left( U_{\| i}  \N^{ 01} \right) - \frac{1}{\rho_*} \poissonbracket{\phi}{\N^{ 01}} \nonumber \\  &  + \C^{01}_i  +  S_\mathcal{E}^{01}.
\end{align}
\end{subequations}
\\
Using \eqref{eq:dn20dtanddn01dt} with the ion continuity equation given in \eqref{eq:nicoldion} into \eqref{eq:dtempidt}, we derive 

\begin{align} \label{eq:nidtempidt}
N_i \partial_t T_i & = - \frac{\sqrt{2\tau_i }}{3}\partial_z \left(   \sqrt{3} \N^{30} -  \sqrt{2} \N^{11}\right) \nonumber \\
&- U_{\| i } N_i \partial_z T_i   - \frac{N_i}{ \rho^*} \poissonbracket{\phi}{T_i} - \frac{2}{3} P_{\parallel i} \partial_z U_{\| i }   \nonumber \\ 
& + S_{\mathcal{E}} + \left( 1 - T_i \right) S_{\N} ,
\end{align}
\\
where the collision terms, $  \C^{20}_i $ and $  \C^{01}_i$, cancel out due to the energy conservation of the collision operator (see \eqref{eq:moments:nonlinear_dougherty}). In addition, the definitions of the sources given in \eqref{eq:sourcesnpj} are used to derive the last terms on the right-hand side of \eqref{eq:nidtempidt}.

We now use the fact that FLR effect are neglected in our model such that $N_i \simeq n_e$. Moreover, an expression for the parallel gradient of the parallel ion flow, $\partial_z U_{\| i}$, can be obtained from the vorticity equation given in \eqref{eq:vorticitybraginskii}. At the leading-order in the long-wavelength limit, one finds that $\partial_z J_{\parallel}  \simeq 0$, such that 

\begin{align} \label{eq:nidzupari}
N_i\partial_z U_{\| i } \simeq \left( U_{\| e }  -U_{\| i }  \right) \partial_z n_e + n_e \partial_z  U_{\| e }.
\end{align}
\\
Finally, introducing the normalized parallel ion heat flux,

\begin{align} \label{eq:qiflux}
q_{\parallel i } =  \sqrt{2 \tau_i} \left( \frac{\sqrt{3}}{2} \N^{30} - \frac{\N^{11}}{\sqrt{2}} \right),
\end{align}
\\
and using \eqref{eq:nidzupari}, the ion temperature equation \eqref{eq:nidtempidt} becomes 

\begin{align} \label{eq:tinoclosed}
\partial_t T_i + \frac{1}{ \rho^*} \poissonbracket{\phi}{T_i} + U_{\| i } \partial_z T_i & =  +\frac{2}{3} T_{\parallel i } \left[\left(U_{\parallel i}- U_{\parallel e}\right) \frac{\partial_z  n_e}{n_e}- \partial_z U_{\parallel e}\right]    \nonumber \\ 
& - \frac{2 }{3} \frac{1}{n_e}\partial_z q_{\parallel i } + \frac{A_{\mathcal{E}}}{n_e} + \left( 1 - T_i \right) \frac{S_{\N} }{n_e}.
\end{align}
\\
We remark that \eqref{eq:tinoclosed} can also be deduced from other full-F gyrofluid models \cite{Held2016,Wiesenberger2022}, which consider finite ion temperature effects with a fixed number of GMs. To close \eqref{eq:tinoclosed}, an expression for the parallel ion heat flux $q_{\parallel i}$ must be prescribed. In contrast to the ion GM hierarchy equation, which treats the ion parallel heat flux $q_{\parallel i}$ as a dynamical variable by evolving high-order GMs (see \eqref{eq:qiflux}), the Braginskii model assumes a high-collisional limit (where $\lambda_{mpf} k_\parallel \ll 1$), allowing for the derivation of a collisional closure for $q_{\parallel i}$. Specifically, using the Fokker-Planck collision operator and applying the high-collisional closure via the Chapman-Enskog procedure \cite{Chapman1941, Braginskii1965, Jorge2017}, we can obtain the fluid closure $q_{\parallel i} \simeq -\chi_{\parallel} \partial_z T_i$, where $\chi_{\parallel}$ represents the normalized parallel ion thermal conductivity predicted by the Braginskii transport equation \cite{Braginskii1965, Zeiler1997}. Hence, in the high-collisional limit, the ion temperature equation \eqref{eq:tinoclosed} reduces to 

\begin{align} \label{eq:tifinal}
\partial_t T_i + \frac{1}{ \rho^*} \poissonbracket{\phi}{T_i} + U_{\| i } \partial_z T_i & =  +\frac{2}{3} T_{i } \left[\left(U_{\parallel i}- U_{\parallel e}\right) \frac{\partial_z  n_e}{n_e}- \partial_z U_{\parallel e}\right]    \nonumber \\ 
& + \partial_z \left( \chi_{\parallel i} \partial_z T_i  \right) + \frac{A_{\mathcal{E}}}{n_e} + \left( 1 - T_i \right) \frac{S_{\N} }{n_e},
\end{align}
\\
where the temperature anisotropy is neglected such that $T_{\parallel i } \simeq T_i$ at high-collisionality \cite{Frei2022}. Hence, \eqref{eq:tifinal} is equivalent to the ion temperature equation given in \eqref{eq:tibraginskii} used in the Braginskii model where the source terms are due to the density and energy sources (see \eqref{eq:sourcesnpj}) considered in the ion GM hierarchy equation. 

A high-collisional closure for the parallel ion heat flux can also be derived using the Dougherty collision operator (or other collision operator models). However, it has been shown that simplified collision operators can lead to large deviations (with respect to the Fokker-Planck collision operator) in the predictions of the conductivities \cite{Frei2022b}, especially when using operators like Dougherty, as in this work.

Finally, it is worth noting that, while the Braginskii collisional closure diverges in the low-collisionality limit \cite{Pitzal2023}, resulting in an overestimate of the parallel heat flux, the GM hierarchy equation allows for the evolution of $q_{\parallel i}$ using a larger number of GMs. However, while this effect may not be as significant in LAPD due to the low temperature, it could be more pronounced in the boundary region.

\bibliography{biblio}

\end{document}